\documentclass[acmsmall]{acmart}

\AtBeginDocument{ \providecommand\BibTeX{{ \normalfont
B\kern-0.5em{\scshape i\kern-0.25em b}\kern-0.8em\TeX}}}

\usepackage{subcaption}
\usepackage{listings}
\usepackage{comment}
\definecolor{darkgreen}{rgb}{0,0.5,0}

\usepackage[noend]{algpseudocode}
\usepackage{algorithm}
\usepackage{tcolorbox}
\usepackage{tikz}
\usetikzlibrary{matrix,chains,positioning,decorations.pathreplacing,arrows}

\newcommand{\commentED}[1]{{\color{brown}[Eddie: #1]}}

\newcommand{\Ignore}[1]{}
\newcommand{\eg}{{\it e.g.}}
\newcommand{\etc}{etc.}
\newcommand{\ann}{ANN}
\newcommand{\bmc}{BMC}

\newcommand{\etal}{{\it et al.}}

\newtheorem{theorem}{Theorem}[section]

\newtheorem{lemma}[theorem]{Lemma}
\newcommand{\fwl}[3]{\mathcal{F}_{\langle #2, #3 \rangle}\left(#1\right)}

\begin{document}

\title{Incremental Verification of Fixed-Point Implementations of Neural Networks}

\author{Luiz Sena} \affiliation{\institution{Federal University of Amazonas}} \email{coelho.luiz.sena@gmail.com}

\author{Erickson Alves} \affiliation{\institution{Sidia Institute of Science and Technology}} \email{erickson.higor@gmail.com}

\author{Iury Bessa} \affiliation{\institution{Federal University of Amazonas}} \email{iurybessa@ufam.edu.br}

\author{Eddie Filho} \affiliation{\institution{TPV}} \email{eddie.filho@tpv-tech.com}

\author{Lucas Cordeiro} \affiliation{\institution{University of Manchester}} \email{lucas.cordeiro@manchester.ac.uk}

\begin{abstract}
Implementations of artificial neural networks (ANNs) might lead to failures, which are hardly predicted in the design phase since ANNs are highly parallel and their parameters are barely interpretable. Here, we develop and evaluate a novel symbolic verification framework using incremental bounded model checking (BMC), satisfiability modulo theories (SMT), and invariant inference, to obtain adversarial cases and validate coverage methods in a multi-layer perceptron (MLP). We exploit incremental BMC based on interval analysis to compute boundaries from a neuron's input. Then, the latter are propagated to effectively find a neuron's output since it is the input of the next one.  This paper describes the first bit-precise symbolic verification framework to reason over actual implementations of ANNs in CUDA, based on invariant inference, therefore providing further guarantees about finite-precision arithmetic and its rounding errors, which are routinely ignored in the existing literature. We have implemented the proposed approach on top of the efficient SMT-based bounded model checker (ESBMC), and its experimental results show that it can successfully verify safety properties, in actual implementations of ANNs, and generate real adversarial cases in MLPs. Our approach was able to verify and produce adversarial examples for 85.8\%  of 21 test cases considering different input images, and 100\% of the properties related to covering methods. Although our verification time is higher than existing approaches, our methodology can consider fixed-point implementation aspects that are disregarded by the state-of-the-art verification methodologies. 
\end{abstract}


\maketitle

\section{Introduction}
\label{sec:intro}

Artificial neural networks (ANNs) are soft computing models usually employed for regression, machine learning, and pattern recognition problems~\citep{bishop2006PRML} . \ann{}s reflect the behavior of biological neural networks, making them a suitable paradigm for learning tasks and have been recently used to perform various safety-critical tasks. For instance, Amato~\etal{} used \ann{}s to predict medical diagnosis~\cite{amato2013artificial}, while Bojarski~\etal~\cite{bojarski2016end} employed them to perform steering commands in self-driving cars. Unfortunately, in such a context, incorrect classifications\Ignore{misclassifications} can cause serious problems. Indeed, adversarial disturbances can make ANNs misclassify objects\Ignore{simple patterns}, thus causing severe damage to users of safety-critical systems. For instance, Eykholt~\etal{}~\cite{DBLP:conf/cvpr/EykholtEF0RXPKS18} showed that noise and disturbances, such as graffiti on traffic signals, could result in target misclassification, during operation of computer vision systems. Moreover, given that ANNs are notorious for being difficult to interpret and debug, the whole scenario becomes even more problematic~\cite{LundbergL17}, which then claims for techniques able to assess their structures and verify results and behaviors.

Prior work, available in the literature, focused on verifying robustness for manually-written \emph{models} of neural networks~\cite{zheng2016improving,katz2017reluplex,huang2017safety, SunWRHKK18}. The typical approach consists in modeling an ANN and its corresponding verification properties, in SMT-LIB~\cite{Barrett10c}, using integer and real arithmetic theories, and then employ off-the-shelf SMT solvers to find property violations. Recently, Murthy~\etal{}~\cite{Murthy2019} have used SMT to quantify neural-network robustness regarding parameter perturbation. Unfortunately, such verification schemes can not precisely capture issues that appear in \emph{implementations} of \ann{}s, for two main reasons: \textit{(i)}~one can not model bit-level operations using the theory of integer and real arithmetic~\cite{CordeiroFM12}, and \textit{(ii)}~libraries, such as TensorFlow~\cite{tensor-flow}, often take advantage of available graphics processing units (GPUs) to explore the inherent parallelism of \ann{}s; the translation to GPUs can be problematic~\cite{pmlr-v97-odena19a,Sena20}.  Finally, it has been pointed out by Odena~\etal{}~\cite{pmlr-v97-odena19a}  that there exist errors in the Tensorflow graph representation of neural networks,  such as numerical errors and disagreements between ANN implementations and their quantized versions. 

We propose a verification technique that exploits \ann{} implementations, through operational models, to infer invariants, which is based on incremental bounded model checking (BMC). BMC is a popular verification technique that translates a program into a formula and then uses Boolean Satisfiability (SAT) or Satisfiability Modulo Theories (SMT) solvers to check bugs. Moreover, that happens up to some bound $k$ on the depth of a program (\eg{}, depth of data structures, loop iterations, call chains, \etc), whose choice is critical for the efficiency of a \bmc{} tool~\cite{GuntherW14}. In particular, the incremental BMC refers to the repeated execution of \bmc{} for increasing values of $k$, until all loops are unrolled, or maximum depth is reached. Here, our main contribution is an \ann-custom verification method for inferring invariants based on intervals, which are introduced into an ANN implementation as assumptions based on incremental BMC. For instance, Sigmoid is an activation function commonly used in MLP~\cite{bishop2006PRML}, which contains one or more hidden layers (apart from one input and one output layer), whose output ranges from $0$ to $1$.
Consequently, this information is added to a verification model, as an assumption to help prune the state-space exploration. One may notice that activation functions used for \ann{}s are nonlinear elements with mathematical operations that are computationally expensive for SAT/SMT solvers. Nonetheless, depending on the weights and the range of input values, these operations can be avoided since the behavior of such nonlinear functions may be constant for some intervals. Therefore, the invariant inference can help reduce verification times by simplifying \ann{}s' output computation for some input intervals, on an incremental BMC framework.

Thus, we exploit incremental BMC based on invariant inference to verify actual implementations of ANNs in CUDA~\cite{10.5555/2935593}. Each inferred invariant is converted into an assumption and propagated through the control flow to each program point; we use these assumptions to encode and check path guards, thereby avoiding the exploration of unfeasible paths. In the proposed symbolic verification framework, boundaries are computed from a neuron's input and propagated to find its effective output, given that the input of a neuron is the output of the previous one. In addition to checking a range of language-specific safety properties, such as the absence of arithmetic under- and overflow, out-of-bounds array indexing, and null-pointer dereferencing, we also consider properties that are inherent to the design of \ann{}s, such as covering-method validation based on modified condition/decision coverage (MC/DC)~\cite{hayhurst2001practical}, which measures how adversarial two images are. We also obtain adversarial cases from ANNs that could lead to system malfunctioning.
Moreover, if real implementations are tackled, one additional problem arises: the word length used to store data and operate over them. Data representation in computers is inherent finite, even with floating-point~\cite{Bessa2016}. It may become a considerable concern when fixed-point arithmetic comes into play, which is usually known as the finite word-length (FWL) problem. The proposed framework takes it into account and can check an ANN implemented with a particular FWL format, which means those safe entities could be promptly deployed on real target systems.

Our approach is implemented on top of the efficient SMT-based bounded model checker  (ESBMC)~\cite{esbmc-fase-2020,GadelhaMCN19}, which is extended with operational  models~\cite{MonteiroGCF17,PereiraASMMFC17} that handle the CUDA deep learning primitives  cuDNN~\cite{chetlur2014cudnn} and cuBLAS~\cite{nvidia2008cublas}, being that a simple and elegant verification approach.  Instead of providing a complex scheme or model capable of capturing properties of neural networks,  we instrument basic calls and verify a program's correctness regarding their use, as implemented on a real platform. A pattern recognition benchmark~\cite{Sena20} is employed to evaluate the performance and correctness of the proposed approach. We have performed conformance testing~\cite{KrichenT09} over our CUDA (operational) models to determine whether they comply with the requirements of the ANNs, which was performed through exhaustive execution with deterministic inputs. This way, traceable behavior was sought to reproduce the same results. 

Our experimental results show that ESBMC correctly validates all covering methods. Verification times were no longer than a minute, when checking how adversarial two images are, concerning the ANN neurons. Additionally,  the proposed framework was able to produce $19$ adversarial cases, which was confirmed by validation scripts in MATLAB. This step is required to concretize and prove that real adversarial case exists. However, the associated verification times were higher than other existing approaches (\eg{}., Deep Learning Verification~\cite{huang2017safety}), due to the bit-accurate precision of the employed verification process. In summary, this paper describes the following additional original contributions:

\begin{itemize}
 \item development of an operational model (OM) to handle CUDA primitives (cf. Section~\ref{ssec:method});
 \item development of a methodology that tackles challenges specific to ANNs, such as coverage, which uses mutants for activating new neurons, adversarial cases, and finite-precision arithmetic and its rounding errors (cf. Sections~\ref{ssec:coverage} and~\ref{ssec:adversarial});
 \item speed-up of one order of magnitude, over a conventional
incremental BMC engine and when checking covering methods, due to the use of invariants (cf. Section~\ref{ssec:results});
\item provision of research artifacts, including benchmarks, tools, and experimental results, which are all available for download.
\end{itemize}

Experimental results show that the proposed approach can efficiently validate coverage methods in a few minutes. Moreover, implementations of ANNs, for a set of intricate benchmarks obtained from the available literature, can be adequately verified: the median run-time for our benchmark set, considering the incremental BMC approach and using invariant inference, is around $46$h. Despite the high verification time, our proposed approach was able to find violations in ANN CUDA implementations, considering the fixed-point arithmetic and FWL, in 85.8 \% of 21 test cases for a vocalic images recognition benchmark.

The remaining of this paper is organized as follows. Section~\ref{sec:preliminaries} provides preliminaries about artificial neural networks, including implementation aspects, activation functions, and notions of bounded model checking of software with particular focus on ANN implementations. Section~\ref{sec:verification} tackles our incremental verification method using invariant inference by focusing on the base case and forward condition steps; we also provide details of our invariant inference based on interval analysis and operational models for ANNs in CUDA. Section~\ref{sec:ValidationVerification} describes the validation of covering methods and our verification algorithm to obtain adversarial cases in ANN implementations. Section~\ref{sec:exp} presents our experimental results for covering methods and adversarial examples. Section~\ref{sec:related_work} analyses the related work. Finally, Section~\ref{sec:conclusion} concludes and outlines future work.

\section{Preliminaries}
\label{sec:preliminaries}

\subsection{Artificial Neural Networks (ANNs)}
\label{ssec:ann}

ANNs are efficient models for statistical pattern recognition, making them a suitable paradigm for learning tasks~\cite{bishop2006PRML}. Learning, in ANNs, is based on modifying the {\it synaptic} weights of their interconnected units, which will fit their values according to a given set of labeled examples or tasks, during a training phase. A training algorithm uses each input example to converge these weights to specific results. Mainly, ANNs are broadly employed to solve classification problems, and one crucial property regarding that is their generalization ability for adversarial situations~\cite{hagan1996neural}, {\it i.e.}, when input patterns are disturbed, for instance, by noise, occlusion, or even sensor defects. Nonetheless, such an ability may not be enough to avoid misclassification, depending on the quality of input features, which may lead to severe problems. For instance, an ANN used in vehicles' driver assistant systems, which was developed for recognizing traffic signs~\cite{Islam2017}, may indicate a wrong sign due to adversarial visibility conditions and then cause accidents.

\begin{figure}[ht]
	\def\layersep{2.5cm}
\resizebox{0.85\linewidth}{!}{
\begin{tikzpicture}[shorten >=1pt,->,draw=black!80, node distance=\layersep]
\tikzstyle{every pin edge}=[<-,shorten <=1pt]
\tikzstyle{neuron}=[circle,draw=black!50,fill=black!25,minimum size=17pt,inner sep=0pt]
\tikzstyle{input neuron}=[neuron, fill=green!30];
\tikzstyle{output neuron}=[neuron, fill=red!30];
\tikzstyle{hidden neuron}=[neuron, fill=blue!30];
\tikzstyle{annot} = [text width=4em, text centered];
\tikzstyle{input}=[rectangle, fill=black];
\foreach \name / \y in {1,...,4}
\node[input, pin=left:{\begin{tabular}{c} Input \#\y \\ $i_{\y}$\end{tabular}}] (IP-\name) at (-0.8*\layersep,-\y) {};
\foreach \name / \y in {1,...,4}
\node[input neuron] (I-\name) at (0,-\y) {$n_{\y,1}$};
\foreach \name / \y in {1,...,5}
\path[yshift=0.5cm]
node[hidden neuron] (H1-\name) at (\layersep,-\y cm) {$n_{\y,2}$};
\foreach \name / \y in {1,...,5}
\path[yshift=0.5cm]
node[hidden neuron] (H2-\name) at (2*\layersep,-\y cm) {$n_{\y,l}$};
\foreach \name / \y in {1,...,5}
\path[yshift=0.5cm]
node[hidden neuron] (H3-\name) at (3*\layersep,-\y cm) {$\dots$};
\foreach \name / \y in {1,...,3}
\node[output neuron,pin=right:{\begin{tabular}{c} Output \#\y \\ $o_{\y}$\end{tabular}}, right of=H3-\y, yshift=-1 cm] (O-\y) {$n_{\y,L}$};
\foreach \source in {1,...,4}
\foreach \dest in {1,...,4}
\path (IP-\source) edge (I-\dest);
\foreach \source in {1,...,4}
\foreach \dest in {1,...,5}
\path (I-\source) edge (H1-\dest);
\foreach \source in {1,...,5}
\foreach \dest in {1,...,5}
\path (H1-\source) edge (H2-\dest);    
\foreach \source in {1,...,5}
\foreach \dest in {1,...,5}
\path (H2-\source) edge (H3-\dest);    
\foreach \source in {1,...,5}
\foreach \dest in {1,...,3}
\path (H3-\source) edge (O-\dest);
\node[annot,above of=H2-1, node distance=1cm] (hl2) {$\dots$};
\node[annot,above of=H1-1, node distance=1cm] (hl1) {Hidden layer {\footnotesize $l=2$}};
\node[annot,above of=H3-1, node distance=1cm] (hl3) {Hidden layer {\footnotesize $l=L-1$}};
\node[annot,left of=hl1] {Input layer {\footnotesize $l=1$}};
\node[annot,right of=hl3] {Output layer {\footnotesize $l=L$}};
\end{tikzpicture}
}
\caption{Example of an ANN.}
\label{fig:annexample}
\end{figure}
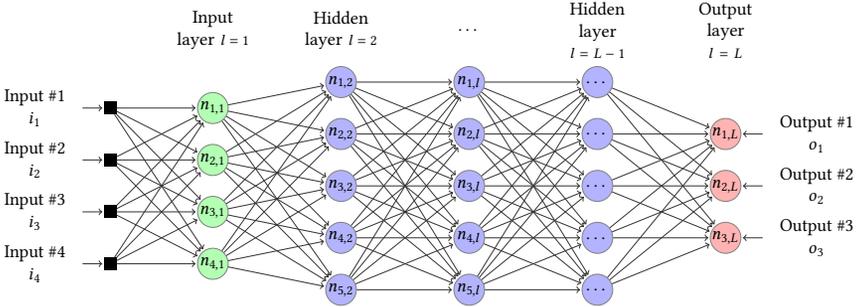

\begin{figure}[ht]
	\resizebox{0.45\linewidth}{!}{
		\begin{tikzpicture}[
		init/.style={
			draw,
			circle,
			inner sep=2pt,
			font=\Huge,
			join = by -latex
		},
		squa/.style={
			draw,
			inner sep=2pt,
			font=\Large,
			join = by -latex
		},
		start chain=2,node distance=13mm
		]
		\node[on chain=2] 
		(x2) {$i_j~$};
		\node[on chain=2,init] (sigma) 
		{$\displaystyle\Sigma$};
		\node[on chain=2,squa,label=above:{\parbox{2cm}{\centering Activate \\ function}}] (actfun)  
		{$\mathcal{N}(u_{k,l})$};
		\node[on chain=2,label=above:Output,join=by -latex] 
		{$y_{k,l}$};
		\begin{scope}[start chain=1]
		\node[on chain=1] at (0,1.5cm) 
		(x1) {$i_1~$};
		\end{scope}
		\begin{scope}[start chain=3]
		\node[on chain=3] at (0,-1.5cm) 
		(x3) {$i_M$};
		\end{scope}
		
		\node[label=above:\parbox{2cm}{\centering Bias \\ $b_{k,l}$}] at (sigma|-x1) (b) {};
		\node[above of = x2, node distance = 0.85cm] (dots) {$\vdots~$};
		\node[below of = x2, node distance = 0.8cm] (dots) {$\vdots~$};
		\draw[o-latex] (x1.east) -- node[above] {$w_{1,k}^{l}$} (sigma);
		\draw[o-latex] (x2.east) + (-0.15,0) -- node[above] {$w_{j,k}^{l}$} (sigma);
		\draw[o-latex] (x3.east) -- node[below] {$w_{M,k}^{l}$} (sigma);
		\draw[o-latex] (b) -- (sigma);
		\draw[-latex] (sigma) -- node[above] {$u_{k,l}$} (actfun);
		\node[below of = x3, xshift = 1.2cm, yshift = 1cm] (weights) {Weights}; 
		\draw[decorate,decoration={brace,mirror}] (x1.north west) -- node[left=10pt] {Inputs} (x3.south west);
		\end{tikzpicture}
	}
	\caption{Detailed view of a single neuron $n_{k,l}$.}
	\label{fig:neuron}
\end{figure}
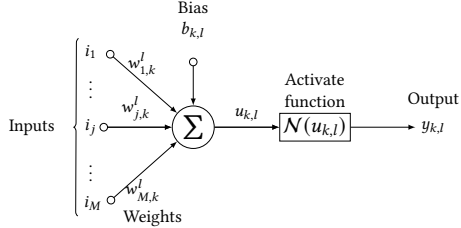

There exist various algorithms to train ANNs: \textit{backpropagation} is a commonly used example~\cite{hagan1996neural}. ANN architectures vary in type of layers, activation functions, number of layers, and neurons. One example is MLP~\cite{haykin2009neural}, which is a classic and commonly used ANN that consists of at least three layers: input, hidden, which employs a non-linear activation function, and output layers. Fig.~\ref{fig:annexample} illustrates an example of ANN with $L$ layers and $N_{l}$ neurons for each $l$-th layer, with $l=1,\dots L$. The ANN has $M=N_{1}=4$ inputs, and $O=N_{5}=3$ output. In this ANN, each neuron of the $l$-th layer, denoted by $n_{k,l}$, with $k=1,\dots N_{l}$, is connected to all neurons $n_{j,l-1}$ of the previous one, with $j=1,\dots N_{l-1}$. Each neuron is structured as shown in Fig.~\ref{fig:neuron}.

The activation potential $u_{k,l}$ of an input \textit{i} = \{$
i_{1}, i_{2}..., i_{M} $\}, for neuron $n_{k,l}$, is computed with
\begin{equation}
u_{k,l}\left(i\right) = \sum_{j=1}^{M}{w_{j,k}^{l} i_{j}} + b_{k,l},
\label{eq:anncalc}
\end{equation}
\noindent where $w_{j,k}^{l}$ is the synaptic weight between
neurons $n_{j,l-1}$ and $n_{k,l}$, and $b_{k,l}$ is the bias of $n_{k,l}$. The neuron output $y_{k,l}$
is computed by means of the activation function
$\mathcal{N}(\cdot) $ evaluated for the activation potential
$u_{k,l}$.
ReLU~\cite{li2017convergence}, Sigmoid,
and Gaussian~\cite{bishop2006PRML} are activation functions
commonly used in MLP. Our experimental results are obtained from MLPs implemented with the activation functions Sigmoid and ReLU. Therefore, the output of $n_{k,l}$ is computed as
\begin{equation}
y_{k,l} = \mathcal{N}(u_{k,l}) =\frac{\mathrm{1} }{\mathrm{1} + e^{-u_{k,l}}}.
\label{eq:neuronout}
\end{equation}

\subsection{Bounded Model Checking of Software}
\label{ssec:bmc}

BMC has been successfully applied to find errors in software systems, including commercial
modules~\cite{CordeiroFM12}. The idea of BMC is to unwind a
program and its correctness properties $k$ times, while generating a
logical formula, in a decidable fragment of first-order logic,
that is satisfiable \textit{iff} a counterexample of size $k$
(or lower) exists~\cite{BiereCCFZ99}. Consequently, this technique is
incomplete, as there might still be a counterexample that
requires more than \textit{k} unwinds to be detected.

As an example, Fig.~\ref{figure:verification-condition} shows
a code fragment that normalizes an image by changing the range
of pixel intensity values.
Figure~\ref{figure:verification-condition}(a) shows the original
code, while Fig.~\ref{figure:verification-condition}(b) shows the
corresponding C program in static single assignment (SSA) form, where every variable has only one definition. This is achieved by introducing a fresh variable from
an original name (\eg{}, with a subscript), such that every
assignment has a unique left-hand side, as shown in
Fig.~\ref{figure:verification-condition}(b). One may notice that BMC
unwinds this program up to a bound $SIZE$ and translates the
corresponding SSA into a verification condition $\psi$, which is satisfiable \textit{iff} the assertion in line $18$ of
Figure~\ref{figure:verification-condition}(a), {\it i.e.}, $img[index] \,
>= \, 0.0f \, \&\& \, img[index] \, <= \, 1.0f$, fails. 
One may notice that standard IEEE 754-2008~\cite{4610935} defines arithmetic operations, conversion and comparison methods, and also total ordering of floating-point numbers~\cite{abs-2004-12699}. Our underlying verification engine follows that standard for reasoning over ANN implementations that contain floating-point numbers~\cite{GadelhaCN17}.

It is also worth mentioning that BMC can non-deterministically choose assignment values for variables while limiting their range with program instructions to make them similar to real user-scenarios. These aspects help prune state-space exploration, which makes BMC less susceptible to the state space explosion problem~\cite{CordeiroFM12}. Both aspects are respectively illustrated in lines $11$ and $12$ of Figure~\ref{figure:verification-condition}(a). Also, SSA representation can be translated into an SMT-LIB problem using either bit-vectors or the abstract numerical domains (e.g., $\mathbb{Z}$, $\mathbb{R}$), while an SMT solver is used to reason about satisfiability, {\it i.e.}, check whether a violation is found.

\begin{figure}[ht]
\centering
\begin{subfigure}{0.4\textwidth}
 \begin{lstlisting}[numbers=left]
#include <assert.h>
#define SIZE 5
typedef unsigned short int ushort;
void normalizef(float *image, ushort size) {
 ushort i;
 for (i = 0; i < size; i++) {
 image[i] = (image[i]) * (1.0f/255.0f);
 }
}
int main(void){
 ushort x1 = nondet_ushort();
 assume((x1 >=0) && (x1 <= 255));
 ...	
 float img[SIZE] = {x1,x2,x3,x4,x5};
 normalizef(img, SIZE);
 uint index = nondet_ushort();
 assume(index>=0 && index <SIZE);
 assert(img[index]>=0.0f && img[index]<=1.0f);
 return 0;
}
\end{lstlisting}
\caption{}
\end{subfigure}
\hspace{0.05\textwidth}
\begin{subfigure}{0.4\textwidth}
 \begin{lstlisting}[numbers=left]  
x1#1 == return_value_nondet_ushort#1
assume x1#1 >= 0 && x1#1 < 255
...
x5#1 == return_value_nondet_ushort#5
assume x5#1 >= 0 && x5#1 < 255
index#1 == return_value_nondet_ushort#6
assume index#1 >= 0 && index#1 < 5
assert img#1[index#1] >= 0.0f && img#1[index#1] <= 1.0f 
\end{lstlisting}
\caption{}
\end{subfigure}
\caption{(a) A simple C program to normalize an image. (b) A simplified version of the
 corresponding unwound C program of (a) converted into SSA
 form, where ``...'' means missing statements omitted for simplicity.} 
\label{figure:verification-condition}
\end{figure}

\subsubsection{Incremental BMC}

Incremental BMC uses an iterative technique and verifies imperative programs for each unwind bound, indefinitely, or until it exhausts time or memory limits. Incremental BMC aims to either find a counterexample up to \textit{k} loop unwindings or unwind all loops fully. This algorithm relies on a symbolic execution engine to increasingly unwind a loop after each iteration. Such an approach is divided into two steps: a search for property violations and a procedure that checks whether all loops were fully unwound. When searching for violations, the incremental-BMC technique replaces all unwinding assertions ({\it i.e.}, assertions to check if a loop was completely unrolled) by unwinding assumptions. Typically, this would lead to unsound behavior; however, the first step can only find property violations, thus reporting that an unwinding assertion failure is not a real bug. The next step checks if all loops in a program were fully unrolled, which is done by ensuring that all the unwinding assertions are unsatisfied. No assertion is checked in the second step because they were already checked in the first one, for the current \textit{k} loop unwinding. 

The incremental BMC algorithm also offers the option to change the increment granularity; the default value is $1$. One may notice that changing the increment value can lead to slower verification times and might not present the shortest possible counterexample for a property violation. Incremental BMC is particularly essential to avoid guessing exceedingly large bounds, which might result in prohibitively significant decision problems, thereby making a verifier to run out of resources, before it can provide a result~\cite{GuntherW14}.

\subsubsection{Verify Actual Implementation of ANNs with Incremental BMC}
\label{sec:VerifyActualImplementationofANNswithIncrementalBMC}

ANNs are typically implemented through matrix operations, such as sum, subtraction, and multiplication, which are particularly expensive for bit-accurate verifiers that check safety for actual implementations of \ann{}s.

CUDA is one of the mainstream programming languages to implement
an ANN. For example, TensorFlow~\cite{tensor-flow} allows
software developers to write ANNs in Python, but it
converts them into CUDA-compatible GPU code using the
\textit{cuBLAS}~\cite{nvidia2008cublas} and
\textit{cuDNN}~\cite{chetlur2014cudnn} libraries. This way, a verification methodology based on actual implementations should directly consider such libraries as basic building blocks to be instrumented, without carrying about the exact model for a neural network. 

Here, we rely on ESBMC~\cite{CordeiroFM12}, which is an award-winning model
checker~\cite{esbmc-fase-2020,GadelhaMCN19} targeted at
verifying real-world C programs. In order to support CUDA
operations, an OM, {\it i.e.}, an abstract
representation of the standard CUDA libraries, which
conservatively approximates their semantics, is employed to verify
CUDA-based programs. This way, verification 
of ANNs and their real implementations are considered in a unified framework.

In particular, ESBMC considers the verification of ANN implementations that are subject to FWL effects, which are further amplified in fixed-point processors~\cite{AbreuGCFS16,ChavesBIFCF18}. Arithmetic operations with fixed-point variable are different from the ones with real numbers, since there are some non-linear phenomenons, {\it e.g.}, overflows and round-offs. Let $\langle I,F \rangle$ denote a fixed-point format and $\mathcal{F}_{\langle I,F \rangle}(x)$ denote a real number $x$ represented in the fixed-point domain, with $I$ bits representing the integer part and $F$ bits the decimal one.  The smallest absolute number $c_m$ that can be represented in such a domain is $c_m=2^{-F}$, and any mathematical operations performed at $\mathcal{F}_{\langle I,F \rangle}(x)$ will introduce errors. Here, we treat fixed-point operators for sums, multiplications, subtractions, and divisions, within our BMC framework, as \texttt{fxp\_add}, \texttt{fxp\_mult}, \texttt{fxp\_sub}, and \texttt{fxp\_div}, respectively.


Lastly, ESBMC also explicitly explores the possible interleavings of C
programs (up to the given context bound), while treating each
interleaving itself symbolically. It employs
monotonic partial order reduction~\cite{KahlonWG09} and
\textit{k}-induction~\cite{esbmc-sttt-2017}, with the goal of efficiently pruning
the state-space exploration, and verifies properties such as
user-specified assertions, deadlocks, memory leaks, invalid
pointer dereference, array out-of-bounds, and division by zero,
in C programs.

\section{Incremental Verification of Neural Networks in CUDA using Invariant Inference}
\label{sec:verification}

\subsection{Motivating Example}

\begin{figure}[htb]
	\centering
	\def\layersep{2.5cm}
	\resizebox{0.5\linewidth}{!}{
		\begin{tikzpicture}[shorten >=1pt,->,draw=black!80, node distance=\layersep]
		\tikzstyle{every pin edge}=[<-,shorten <=1pt]
		\tikzstyle{neuron}=[circle,draw=black!50,fill=black!25,minimum size=17pt,inner sep=0pt];
		\tikzstyle{input neuron}=[neuron, fill=green!30];
		\tikzstyle{output neuron}=[neuron, fill=red!30];
		\tikzstyle{hidden neuron}=[neuron, fill=blue!30];
		\tikzstyle{annot} = [text width=4em, text centered];
		\tikzstyle{input}=[rectangle, fill=black];
		\node[input, pin=left:{\begin{tabular}{c} Input {\#}1 \\ $x$\end{tabular}}] (IP-1) at (-\layersep,-2) {};
		\node[input, pin=left:{\begin{tabular}{c} Input {\#}2 \\ $y$\end{tabular}}] (IP-2) at (-\layersep,-4) {};
		\node[input neuron] (I-1) at (0,-2) {$A$};
		\node[input neuron] (I-2) at (0,-4) {$B$};
		\node[output neuron,pin={[pin edge={->}]right:Output}, right of=I-1, yshift = -1cm] (O) {$f$};
		\path (IP-1) edge node[above,at start, anchor= south west] {2} (I-1);
		\path (IP-2) edge node[above,at start, anchor= south] {-3} (I-1);
		\path (IP-1) edge node[above,at start, anchor= north] {1} (I-2);
		\path (IP-2) edge node[above,at start, anchor= north west] {4} (I-2);
		\path (I-1) edge node[above] {1} (O);
		\path (I-2) edge node[below] {1} (O);
		\end{tikzpicture}
	}
	%
	\caption{Simple fully-connected neural network.}
	\label{fig:motivating-example}
\end{figure}
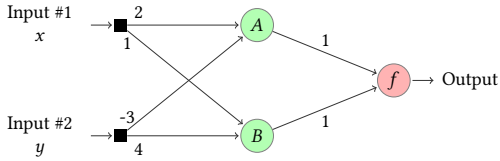

To provide some insight regarding the need for verifying ANN implementations, a simple example is presented. Fig.~\ref{fig:motivating-example} illustrates a simple fully-connected ANN. This example has $2$ inputs, $1$ output, and no biases. From Fig.~\ref{fig:motivating-example}, the ANN's output is directly computed with
\begin{equation}\label{eq:motivating-example}
f = A + B = ReLU(2x - 3y) + ReLU(x + 4y).
\end{equation}

In our approach, we need three particular inputs: ($1$) ANN written in the \texttt{nnet} file format~\cite{nnet-format}; ($2$) input intervals to guide the exploration; and ($3$) safety properties for which we want to validate a given ANN. 
%
%
Let us assume we want to verify if the ANN in Fig.~\ref{fig:motivating-example} is safe concerning property $\phi$, assuming $x,y\in  [0.749, 0.498]$:
\begin{equation}\label{eq:motivating-example-property}
\phi \leftrightarrow f \geq 2.7.
\end{equation}

\noindent First, we will parse the given ANN in the \texttt{nnet} file format to our internal network model. For the sake of readability, we have omitted this translation, since it is fairly straightforward. Second, we will perform the ANN concretization, layer by layer, using our CUDA operational model (mainly the \texttt{cublasSgemm} function), as well as taking the activation function of each neuron into account. The activation function for the hidden layer's neurons is $ReLU : {\rm I\!R} \rightarrow {\rm I\!R}^+$~\cite{li2017convergence}, {\it i.e.},
\begin{equation}
ReLU(x) = max(0, x).
\end{equation}

\noindent Consequently, it returns $0$, for negative inputs, and the input itself, otherwise. Thus, we have:
\begin{equation}
A = ReLU(2 \times 0.749- 3 \times 0.498) = ReLU(0.004) = 0.004,
\end{equation}
%
%
\begin{equation}
B = ReLU(0.749 + 4 \times 0.498) = ReLU(2.741) = 2.741,
\end{equation}
%
\begin{equation}
f = A + B = 0.004 + 2.741 = 2.745,
\end{equation}
\noindent which is used to perform the last step of our approach, \textit{i.e.}, checking the user-defined property $\phi$. In this case, $\phi$ holds for the given input, since $f = 2.745 \geq 2.7$.

At first glance, we can conclude that it is safe, which usually happens in many verification frameworks~\cite{Wang2018}, when a final implementation on a real platform with restrictions is not considered. Nonetheless, if that is the case, the FWL effects should be checked. The performed steps are necessarily the same as in the floating-point representation, except that we use a different version of our CUDA operational model that uses fixed-point arithmetic.

In this work, we have encoded non-integral numbers in a binary representation. Given a rational number, we are able to represent it in fixed-point by using $m + n$ bits, where the integral part $I$ is encoded in $m$ bits and the fractional part $F$ is encoded in $n$ bits. Moreover, it is interpreted as $I + \frac{F}{2^n}$. In our operational model, we have defined conversion functions to represent floating-point values in fixed-point representation and vice-versa, \textit{i.e.}, \texttt{fxp\_to\_float} and \texttt{fxp\_float\_to\_fxp}.

Furthermore, regarding the ANN in Fig.~\ref{fig:motivating-example} and the same property $\phi$ mentioned earlier, while using a representation with $4$ bits for the integral part and $6$ bits for the fractional one, with the operators defined in Section~\ref{sec:VerifyActualImplementationofANNswithIncrementalBMC}, we have $f=\fwl{y_{1,2}}{3}{6}=2.6867$, which violates $\phi$, since $f = 2.687 < 2.7$. Although the example above is straightforward, it reveals the importance of checking ANNs while taking their final implementations into account. One developer could indeed take this ANN, which would be regarded as safe, and only change the number of bits employed for data storage and a suitable arithmetic logic unit, believing that would cause no harm. However, as already seen, the number of bits matters. It is worth mentioning that this problem is even worse for larger ANNs, due to cumulative error, which could lead to very different verification results compared with verification procedures that consider only floating-point arithmetic.

\subsection{Operational Models for ANNs in CUDA}
\label{ssec:method}


Our incremental verification method is based on two phases: (1)
obtain the required models from real ANN programs written in
CUDA and (2) design safety properties that ensure the reliability of
ANN implementations. All models are collected from actual
implementations of ANNs written in CUDA. There exist two APIs in
CUDA to support neural networks and mathematical operations:
\textit{cuBLAS}~\cite{nvidia2008cublas} and
\textit{cuDNN}~\cite{chetlur2014cudnn}. The former provides
mathematical operations, such as matrix multiplication and sums, which
are typically used on the feed-forward process of
neural networks~\cite{hagan1996neural}. The latter, in turn, provides
deep neural networks (DNN) primitives such as tensors
operations, convolution functions, activation functions, and
backward operations~\cite{hagan1996neural}.

\begin{figure}[htb]
\centering
\includegraphics[width=0.7\textwidth]{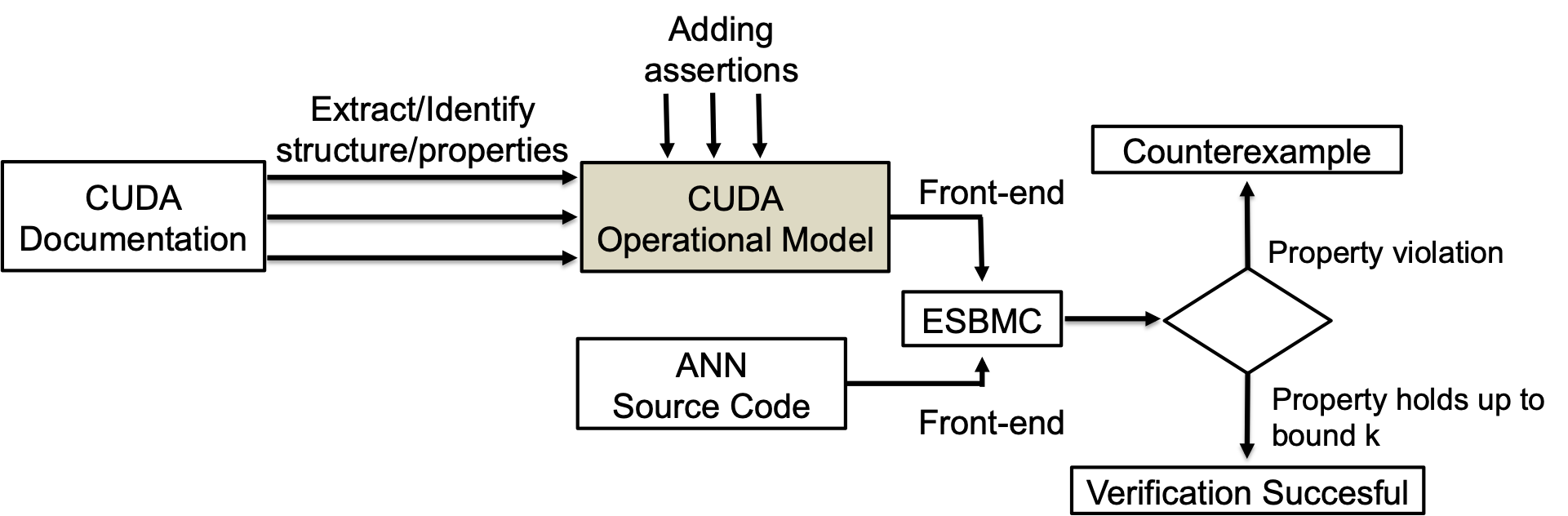}
\caption{CUDA operational model.}
\label{fig:cuda-om}
\end{figure}

Similar to Pereira \etal{}~\cite{PereiraASMMFC17}, we have built our operational model based on the official CUDA documentation~\cite{nvidia2008cublas,chetlur2014cudnn}. First, we have identified the \textit{cuBLAS} and \textit{cuDNN}'s structures and then extracted their pre- and post-conditions. Second, we have written the CUDA code to represent each method's actual behavior implemented in our CUDA OM. Fig.~\ref{fig:cuda-om} illustrates how the CUDA OM (including \textit{cuBLAS} and \textit{cuDNN}) was built and integrated into our verification framework. It is worth noticing that the mentioned OMs are not exact reproductions of the original libraries. Indeed, they aim at following the same logic and output \cite{MonteiroGCF17}, but subsequent operations that do not present resources to be checked are not performed. Consequently, we obtain a simplified and suitable infrastructure to guide the underlying model checker during verification procedures.

Our proposed operational model supports all necessary functions for the feedforward process of MLPs, \eg{}, matrix multiplication functions, and activation functions are all functions provided by cuBLAS and cuDNN.

One may also notice that any ANN written in CUDA and using these APIs can be verified by our method. In practice, a real program would then use the developed OM when calling the mentioned APIs so that the underlying model checker would be able to find property violations. Additionally, any user-specified assertion can be provided and then verified by ESBMC. Our operational models aim to follow the same steps performed in the original libraries, but ignoring irrelevant calls (\eg{}, screen-printing methods), where there is no safety property to be checked.

For the \textit{cuBLAS} and \textit{cuDNN} libraries, we have ensured that their operational models return the same results as the original ones. Indeed, OM methods were exhaustively verified and compared with the original ones, using various deterministic data sets and performing conformance tests~\cite{KrichenT09}. As an example, we can see a pseudo-code of the \textit{cublasSgem} operational model using our fixed-point library (cf. Section~\ref{sec:VerifyActualImplementationofANNswithIncrementalBMC}), in Algorithm~\ref{algo:cublasSgemm}, which consists  of multiplying matrices $A$ and $B$ and storing its result in matrix $C$.

\begin{algorithm}
\scriptsize
\caption{\textit{cublasSgemm}}
\begin{algorithmic}[1]
\Require matrices $A$ and $B$.
\State $\textit{x} \gets \text{ 0 }$
\For {$x < k $}
\State $\textit{y} \gets \text{ 0 }$
\For {$y < i $}
\State $\textit{z} \gets \text{ 0 }$
\State $\textit{sum} \gets \text{ 0 }$
\For {$z < j $}
\State{$sum \gets fxp\_sum(fxp\_mul(A[x][z] * B[z][y]) + sum$)}
\State{$z++$}
\EndFor
\State{C[x][y] = sum}
\State{$y++$}
\EndFor
\State{$x++$}
\EndFor
\Ensure matrix $C$.
\end{algorithmic}
\label{algo:cublasSgemm}
\end{algorithm}

Another important OM in the \textit{cuDNN} API is \texttt{cudnnActivationForward}. In particular, this function operates in different ways by applying a specified neuron activation function element-wise over each input value. We support two activation functions: Sigmoid and RELU. For example, the function Sigmoid described in~\eqref{eq:neuronout} is internally implemented in \texttt{cudnnActivationForward}, which is represented by Algorithm~\ref{algo:cudnnActivationForward} within our OM, using fixed-point arithmetic.
\begin{algorithm}
\scriptsize
\caption{\textit{cudnnActivationForward}}
\begin{algorithmic}[1]
\Require Activation descriptor \textit{activationDesc}.
\Require Pointers \textit{alpha} and \textit{beta} to scaling factors (in host memory) used to blend the computation result with prior value in the output layer.
\If {$activationDesc == CUDNN\_ACTIVATION\_SIGMOID$}
\State $i \gets \text{ 0 }$
\For {$i < layerSize $}
\State{$output[i] \gets sigmoid(fxp\_mul(input[i]*alpha))$}
\State{$output[i] \gets fxp\_mul(output[i]*beta$)}
\State{$i++$}
\EndFor
\State \Return CUDNN\_STATUS\_SUCCESS
\EndIf
\Ensure Data pointer to GPU memory associated with the output tensor descriptor \textit{output}.
\end{algorithmic}
\label{algo:cudnnActivationForward}
\end{algorithm}

\subsection{Incremental Verification using Invariant Inference}
\label{sec:IncrementalVerificationusingInvariantInference}

We develop an incremental verification algorithm to verify and
falsify safety properties in CUDA programs~\cite{GadelhaMCN19}.
Let a given CUDA program $P$, which implements an ANN, be
modeled as a finite transition system $M$, which is defined as
follows:
\begin{itemize}
\item[--] $I(s_n)$ and $T(s_n, s_{n+1})$ as the formulae over
a program's state variable set $s_i$ constraining the initial
states and transition relations of $M$;

\item[--] $\phi(s)$ as the formula encoding states satisfying a
safety property to verify language-specific
properties, validate covering methods (cf.
~\eqref{eq:sscover}--\eqref{eq:dvcover}), and obtain
adversarial cases (cf.~\eqref{eq:imagemisclassified});

\item[--] $\psi(s)$ as the formula encoding states satisfying
the completeness threshold, {\it i.e.}. $\psi(s)$ will contain unwindings no deeper than the maximum number of loop-iterations.
\end{itemize}

One may notice that, in our notation, termination and error are mutually
exclusive: $\phi(s) \wedge \psi(s)$ is
unsatisfiable by construction; $s$ is a deadlock state if $T(s, s') \vee
\phi(s)$ is unsatisfiable. In each step $k$ of the incremental
verification algorithm, two propositions are checked: the base case
$B(k)$ and the forward condition $F(k)$. $B(k)$ represents the
standard BMC and is satisfiable \textit{iff} $P$ has a
counterexample of length \textit{k} or less, {\it i.e.},
\begin{equation}\label{eq:bk}
 B(k) \Leftrightarrow I(s_1) \wedge \left( \bigwedge^{k-1}_{i=1} T(s_i, s_{i+1}) \right) \wedge
\left( \bigvee^{k}_{i=1} \neg \phi(s_i) \right) .
\end{equation}

The forward condition checks for termination, {\it i.e.}, whether the
completeness threshold $\psi(s)$ must hold for the current $k$.
If $F(k)$ is unsatisfiable, $P$ has terminated:
\begin{equation}\label{eq:fk}
 F(k) \Leftrightarrow I(s_1) \wedge \left( \bigwedge^{k-1}_{i=1} T(s_i, s_{i+1}) \right) \wedge \neg
\psi(s_k).
\end{equation}

No safety property $\phi(s)$ is checked in $F(k)$, as they are
already checked for the base case. Finally, the inductive
condition $S(k)$ is unsatisfiable if, whenever $\phi(s)$ holds for
$k$ unwindings, it also holds for the next unwinding of $P$:
\begin{equation}\label{eq:sk}
 S(k) \Leftrightarrow \exists n \in \mathbb{N}^{+} . \bigwedge^{n+k-1}_{i=n} (\phi(s_{i}) \wedge T'(s_i, s_{i+1})) \wedge \neg \phi(s_{n+k}).
\end{equation}
Here, $T'(s_i, s_{i+1})$ is the transition relation after havocking the loop variables~\cite{esbmc-stt-2017}. Through $B(k)$, $F(k)$, $S(k)$, and $\pi(k) \Leftrightarrow B(k) \vee [ F(k) \wedge S(k) ] $, the incremental verification algorithm $\mathrm{bmc}_{\mathrm{inc}}$ to
falsify or verify ANN implementations, at a given $k$, is given by
\begin{equation}\label{eq:verk}
 \mathrm{bmc}_{\mathrm{inc}}(P, k) =
 \begin{cases}
 P \text{ is unsafe}, & \text{if}\ B(k)\ \text{is satisfiable},\\
 P \text{ is safe}, & \text{if}\ \pi(k) \ \text{is unsatisfiable},\\
 \mathrm{bmc}_{\mathrm{inc}}(P, k+1), & \text{otherwise}.
 \end{cases}
\end{equation}

\subsubsection{Invariant Inference based on Interval Analysis}
\label{sssec:invariant}

We employ an interval invariant generator for integer and
floating-point variables within our incremental verification
approach. This invariant generator computes, for every program
variable, a lower and an upper bound of possible values. Those
intervals are injected into the input program as assumptions
(constraints). Similarly to Rocha \etal{}~\cite{RochaRIC017}, we
perform static analysis before loop unwinding and
(over-)estimate the range that a variable can assume. In
contrast to Rocha \etal{}, we do not rely on external tools to
infer polyhedral constraints ({\it e.g.}, $ax + by \leq c$, where $a$,
$b$, and $c$ are constants and $x$ and $y$ are variables) over C
programs. Instead, we implement a ``rectangular'' invariant
generation based on interval analysis ({\it e.g.} $a \leq x \leq b$)
as a pre-processing step of the verification, {\it i.e.}, before an
ANN implementation is symbolically executed, and then the resulting
formulae are checked by an SMT solver.

Here, we use the abstract-interpretation component available in
CPROVER~\cite{cprover-manual} to obtain an abstract
domain based on expressions over intervals; these constraints
associate each variable with upper and lower bounds. This
algorithm begins by assuming an unbounded interval for each
program variable and follows the reachable instructions
from  function \texttt{main}, while updating intervals,
thus merging them, if necessary. When loops are found, a
widening operation is applied to accelerate the
invariant generation process~\cite{YamaguchiBRIK19}. As a result, we
generate new invariants $\varphi(s_i)$ and change
\eqref{eq:bk} and~\eqref{eq:fk} to use them as assumptions
during verification, such that the new $B'(k)$, $F'(k)$, and $S'(k)$ are defined as
\begin{align}
	&  B'(k) \Leftrightarrow I(s_1) \wedge \left( \bigwedge^{k-1}_{i=1} T(s_i, s_{i+1}) \right) \wedge \varphi(s_i) \wedge
	\left( \bigvee^{k}_{i=1} \neg \phi(s_i)\right), \label{eq:newbk} \\
	&  F'(k) \Leftrightarrow I(s_1) \wedge \left(\bigwedge^{k-1}_{i=1} T(s_i, s_{i+1})\right) \wedge \varphi(s_i) \wedge \neg
	\psi(s_k), \label{eq:newfk} \\
	&  S'(k) \Leftrightarrow \exists n \in \mathbb{N}^{+} . \ \varphi(s_n) \wedge \left( \bigwedge^{n+k-1}_{i=n} \phi(s_{i}) \wedge T'(s_i, s_{i+1})\right) \wedge \neg \phi(s_{n+k}). \label{eq:newsk}
\end{align}

As an illustrative example, one should consider our function Sigmoid shown
in Algorithm~\ref{algo:sigmoid}, which is based on a lookup
table. Functions Sigmoid are typically implemented through functions
\texttt{pow} and \texttt{sqrt} (cf.~\eqref{eq:neuronout}), which compute the
value of base raised to the power exponent and the square root,
respectively; these functions are computationally expensive to
calculate. For example, SAT solvers do not scale well when
reasoning on the propositional encoding of arithmetic operators
({\it e.g.}, multiplication and division), because the operands are
treated as arrays of $c$ (where $c$ represents the bit-width of
the data type) unrelated propositional variables; consequently, the computational effort is wasted during the propositional
satisfiability search~\cite{CordeiroFM12}. Additionally, since
the output of function Sigmoid ranges from $0$ to $1$, a lookup
table can represent the sigmoid behaviour with a relatively good
precision. Our lookup table maps inputs between $-20$ and $20$
to values between $0$ and $1$, with a $2$-decimal places
precision. For any value less than $-20$, its output will be $0$
and, for any value greater than $20$, it will be
$1$~\cite{bishop2006PRML}. Such truncations are reasonable since
$\mathcal{N}(20) = 1 - \mathcal{N}(-20) \approx 2\times 10^{-9}
<< 10^{-2}$, {\it i.e.}, the rounding effect is irrelevant, if compared
to $10^{-2}$.

\begin{minipage}{0.45\textwidth}
\begin{algorithm}[H]
	\scriptsize
	\caption{\textit{sigmoidLUT}}
	\begin{algorithmic}[1]
		\Require input signal \textit{u}.
		\State $\textit{index} \gets \text{u * 100 + 2000}$
		\If {$index < 0$}
		\State $\textit{output\_value} \gets \text{0}$
		\Else
		\If {$index \geq 4000$}
		\State $\textit{output\_value} \gets \text{1}$
		\Else
		\State $\textit{output\_value} \gets \text{lookup[index]}$  
		\EndIf
		\EndIf
		\Ensure Output signal \textit{outval}.
	\end{algorithmic}
	\label{algo:sigmoid}
\end{algorithm}
\end{minipage}
\hfill
\begin{minipage}{0.49\textwidth}
\begin{algorithm}[H]
	\scriptsize
	\caption{\textit{sigmoidLUT} with invariants as assumptions}
	\begin{algorithmic}[1]
		\Require input signal \textit{u}.
		\State $\textit{index} \gets \text{u * 100 + 2000}$
		\If {$index < 0$}
		\State \textit{ASSUME} $index \leq -1$
		\State $\textit{outval} \gets \text{0}$
		\Else
		\If {$index \geq 4000$}
		\State \textit{ASSUME} $4001 \leq index$
		\State $\textit{outval} \gets \text{1}$
		\Else
		\State \textit{ASSUME} $ 0 \leq index < 4000$
		\State \textit{ASSUME} $0 \leq lookup[index] \leq 1$
		\State $\textit{outval} \gets \text{lookup[index]}$  
		\EndIf
		\EndIf
		\Ensure Output signal \textit{outval}.
	\end{algorithmic}
	\label{algo:sigmoidinv}
\end{algorithm}
\end{minipage}

One may notice that since the input of a neuron is the output of the
previous one, we employ interval analysis to compute boundaries
(or properties) of the input and propagate them
to support our incremental verification in efficiently and
effectively finding a neuron's output. Our new function Sigmoid,
with invariants as assumptions, is shown in
Algorithm~\ref{algo:sigmoidinv}.

\section{Verification and Validation of Artificial Neural Networks}
\label{sec:ValidationVerification}
This paper introduces the first bit-precise symbolic verification framework suitable to ANNs implemented in CUDA, while taking into account FWL effects, which means that final implementations on target platforms can be directly validated. Moreover, a mutant-based approach was developed in order to activate a larger number of neurons and then find adversarial cases.

\subsection{Validation of Covering Methods}
\label{ssec:coverage}

\begin{figure}[htb]
\centering
\includegraphics[width=0.7\textwidth]{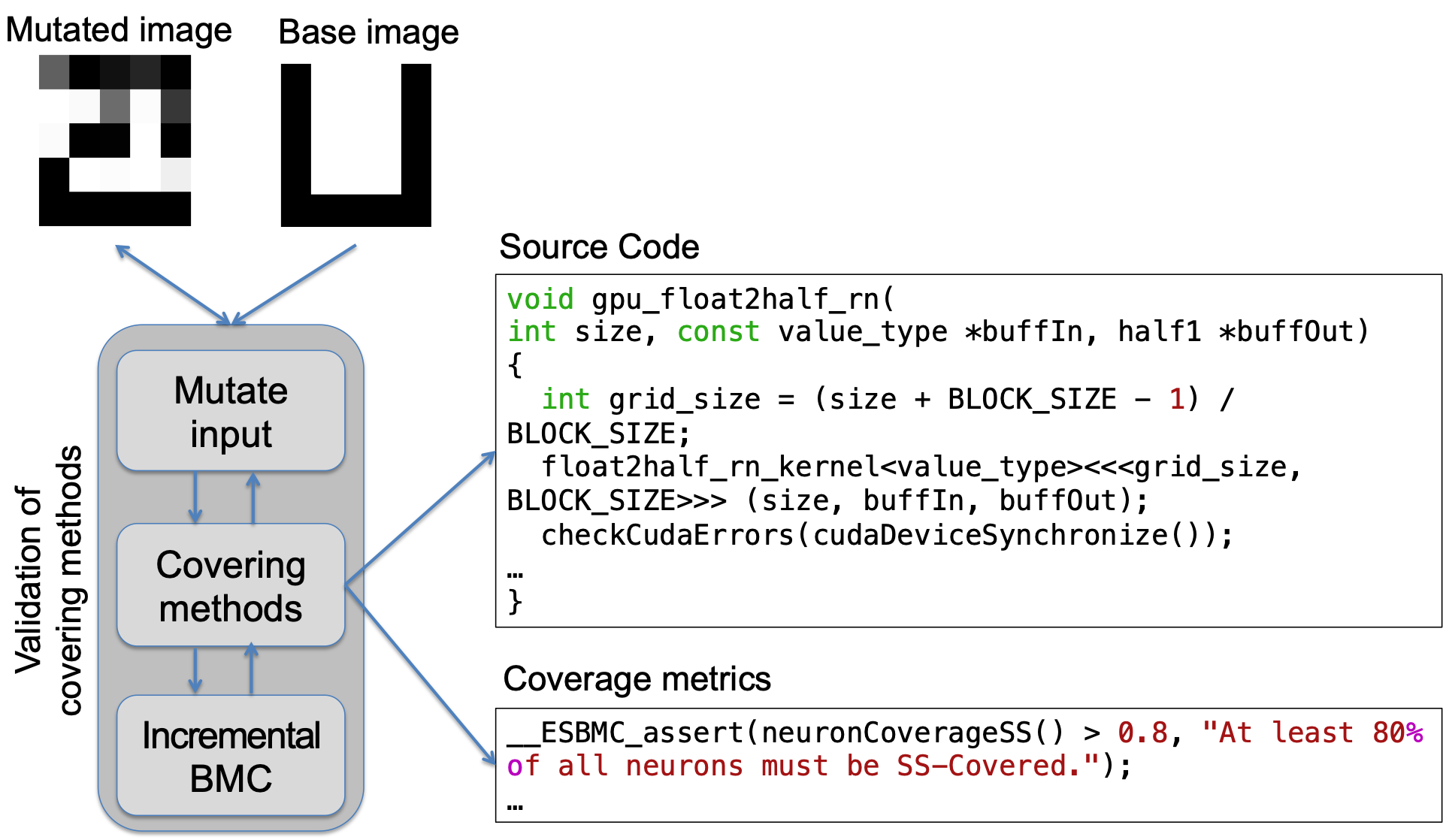}
\caption{Validation of Covering Methods.}
\label{fig:covering-methods}
\end{figure}

Covering methods~\cite{SunWRHKK18} are based on Modified Condition/Decision Coverage (MC/DC)~\cite{hayhurst2001practical}, which is a method applied to ensure adequate testing for safety-critical software. In our symbolic verification framework, MC/DC represents conditions and decisions of an ANN implementation. In particular, conditions are neurons in the previous layer, while decisions are neurons of the following one. We apply covering methods to measure how adversarial two images are, concerning ANN neurons. Fig.~\ref{fig:covering-methods} illustrates an overview of our validation of covering methods. In contrast to Sun \etal{}~\cite{SunWRHKK18}, we exploit incremental BMC to mutate the input and check whether we can achieve a higher neuron-coverage. Indeed, our main challenge lies in generating executions of an ANN implementation that lead to neuron activation.

There exist four covering methods available in the literature~\cite{SunWRHKK18}: \textit{Sign-Sign Cover} (SS-Cover), \textit{Distance-Sign Cover} (DS-Cover), \textit{Sign-Value Cover} (SV-Cover), and \textit{Distance-Value Cover} (DV-Cover), and each of them is implemented and also represented as a property (assertion) within our verification framework. Each property specifies that an image set should lead to a minimum expected coverage regarding all neurons present in an ANN, {\it i.e.}, our algorithm uses each covering method to evaluate whether the adversarial behavior of a pair of images reaches a certain neuron percentage.

In covering methods, neurons are indexed as $n_{k,l}$, where $k$ represents its order on layer $l$, and the considered inputs are defined as $w_{1}$ and $w_{2}$, which are vector inputs (not neuron inputs) representing a pair of elements (in the present case, a pair of images), {\it i.e.}, $w_{1}=\{x_{1,1},x_{2,1}\dots x_{M,1}\}$ and $w_{2}=\{x_{1,2},x_{2,2}\dots x_{M,2}\}$, where $x_{m,p}$ is the ANN input $m$ for element $p$, with $m=\{1,2\dots M\}$ and $p=\{1,2\}$. A signal change, denoted as \textit{sc}, occurs when the activation potential of a certain neuron has its function {\it sign} changed by two different inputs. The function {\it sign} is described as
%
\begin{equation}
sign(i) = \begin{cases}
\text{1,} &\quad\text{if i}\ge0,\\
\text{0,} &\quad\text{otherwise.} \\
\end{cases}
\label{cas:sign}
\end{equation}

\noindent Consequently, $sc (n_{k,l},w_{1},w_{2})$ is \texttt{true} \textit{iff} the activation potential $u_{k,l}$ of neuron $n_{k,l}$ has its signals changed w.r.t. $w_{1}$ and $w_{2}$, {\it i.e.}, $\mathrm{sign}\left(u_{k,l}(w_{1})\right) \neq sign\left(u_{k,l}(w_{2})\right) $. A value change or \textit{vc} occurs when the activation potential of a determined neuron represents a certain value change w.r.t. some metric $g$ and no signal change has occurred, {\it i.e.}, $vc (g,n_{k,l},w_{1},w_{2})$ is {\it true} if $g(u_{k,l}(w_1),u_{k,l}(w_2))$ is \texttt{true} and $\neg sc (n_{k,l},w_{1},w_{2})$, where $g$ is a rational function, \eg{}, $g(a, b) \leftrightarrow \frac{a}{b} \ge d$, and $d$ represents a distance threshold real number that limits the change value. A distance change or \textit{dc}, in turn, occurs when all neurons contained in a certain layer have no signal change and their values represent some value change. Then, $dc (h,l,w_{1},w_{2})$ is defined as \texttt{true} \textit{iff} $\neg sc (n_{k,l},w_{1},w_{2})$ and $h(u_{k,l}(w_1),u_{k,l}(w_2))$ are \texttt{true}, for all neurons in layer $l$. The function $h$, $g$ and $d$ can be any bounded norm-based distance function with any real number that upper bounds it. Here, we define $h$ as $h(u_{k,l}(w_1),u_{k,l}(w_2)) \leftrightarrow dist \left( u_{k,l}(w_1),u_{k,l}(w_2) \right) > v $, where $v$ is a distance threshold. A neuron pair $\alpha$ is denoted as $\alpha = \left(n_{i,l}, n_{k,l+1}\right)$. The covering methods are defined as follows: 
\begin{center}
\begin{tabular}{ll}
	\textbf{\textit{SS-Cover}}: & $ss\left(\alpha, w_{1}, w_{2}\right) = sc\left(n_{i,l},w_{1},w_{2}\right) \wedge sc\left(n_{k,l+1},w_{1},w_{2}\right) \wedge \neg{sc}\left(n_{j,l},w_{1},w_{2}\right) \forall j\neq i$; \\
	\textbf{\textit{SV-Cover}}: & $sv\left(\alpha, g, w_{1}, w_{2}\right) = sc\left(n_{i,l},w_{1},w_{2}\right) \wedge vc\left(g,n_{k,l+1},w_{1},w_{2}\right)\wedge \neg{sc}\left(n_{j,l},w_{1},w_{2}\right) \forall j\neq i$; \\
	\textbf{\textit{DS-Cover}}: & $ds\left(n_{k,l+1},l,h, w_{1}, w_{2}\right) = dc\left(h,l,w_{1},w_{2}\right) \wedge sc\left(n_{k,l+1},w_{1},w_{2}\right)$; \\
	\textbf{\textit{DV-Cover}}: & $dv\left(n_{k,l+1},l,g,h, w_{1}, w_{2}\right) = dc\left(h,l,w_{1},w_{2}\right) \wedge vc\left(g,n_{k,l+1},w_{1},w_{2}\right)$. 
\end{tabular}
\end{center}

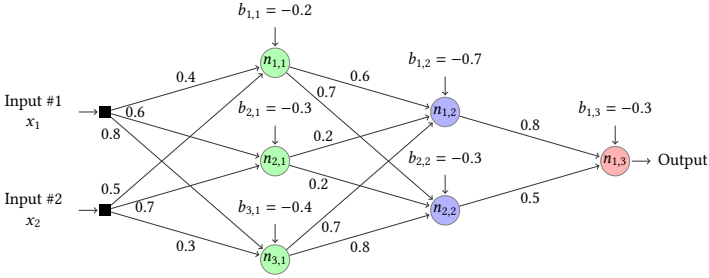
\begin{figure}[htb]
	\centering
	\def\layersep{3.5cm}
	\resizebox{0.7\linewidth}{!}{
		\begin{tikzpicture}[shorten >=1pt,->,draw=black!80, node distance=\layersep]
		\tikzstyle{every pin edge}=[<-,shorten <=1pt]
		\tikzstyle{neuron}=[circle,draw=black!50,fill=black!25,minimum size=17pt,inner sep=0pt]
		\tikzstyle{input neuron}=[neuron, fill=green!30];
		\tikzstyle{output neuron}=[neuron, fill=red!30];
		\tikzstyle{hidden neuron}=[neuron, fill=blue!30];
		\tikzstyle{annot} = [text width=4em, text centered];
		\tikzstyle{input}=[rectangle, fill=black];
		\node[input, pin=left:{\begin{tabular}{c} Input \#1 \\ $x_{1}$\end{tabular}}] (IP-1) at (0,-1) {};
		\node[input, pin=left:{\begin{tabular}{c} Input \#2 \\ $x_{2}$\end{tabular}}] (IP-2) at (0,-3) {};		
		\node[input neuron, pin=above:{$b_{1,1}=-0.2$}] (I-1) at (\layersep,-0 cm) {$n_{1,1}$};
		\node[input neuron, pin=above:{$b_{2,1}=-0.3$}] (I-2) at (\layersep,-2 cm) {$n_{2,1}$};
		\node[input neuron, pin=above:{$b_{3,1}=-0.4$}] (I-3) at (\layersep,-4 cm) {$n_{3,1}$};
		\node[hidden neuron, pin=above:{$b_{1,2}=-0.7$}] (H1-1) at (2*\layersep,-1 cm) {$n_{1,2}$};
		\node[hidden neuron, pin=above:{$b_{2,2}=-0.3$}] (H1-2) at (2*\layersep,-3 cm) {$n_{2,2}$};
		\node[output neuron,pin=above:{$b_{1,3}=-0.3$},pin={[pin edge={->}]right:Output}] (O) at (3*\layersep,-2 cm) {$n_{1,3}$};
		\path (IP-1) edge node[above, anchor= south] {0.4} (I-1);
		\path (IP-2) edge node[above,at start, anchor= south, yshift=0.1cm] {0.5} (I-1);
		\path (IP-1) edge node[above,at start, xshift = 0.5cm, yshift = -0.2cm] {0.6} (I-2);
		\path (IP-2) edge node[above,at start,xshift=0.7cm,yshift=-0.2cm] {0.7} (I-2);
		\path (IP-1) edge node[above,at start, anchor= north, yshift=-0.1cm] {0.8} (I-3);
		\path (IP-2) edge node[above,anchor= north] {0.3} (I-3);
		\path (I-1) edge node[above,anchor= south] {0.6} (H1-1);
		\path (I-1) edge node[above,at start, anchor= west, xshift=0.5cm, yshift=-0.4cm] {0.7} (H1-2);
		\path (I-2) edge node[above,at start,xshift=0.7cm,yshift=0.2cm] {0.2} (H1-1);
		\path (I-2) edge node[above,at start,xshift=0.6cm,yshift=-0.6cm] {0.2} (H1-2);
		\path (I-3) edge node[above,at start, anchor= north west, xshift=0.6cm, yshift=0.7cm] {0.7} (H1-1);
		\path (I-3) edge node[above,anchor= north] {0.8} (H1-2);
		\path (H1-1) edge node[above] {0.8} (O);
		\path (H1-2) edge node[below] {0.5} (O);
\end{tikzpicture}
	}
	\caption{ANN instantiated.}
	\label{fig:netinst}
\end{figure}

Some examples of covering methods can be evaluated with vector inputs (Ex$n$) and neuron outputs available in Table~\ref{tab:cov}, w.r.t. the ANN illustrated in Fig.~\ref{fig:netinst}, which uses the function {\it sigmoid} (cf. Section~\ref{ssec:ann}). Table~\ref{tab:cov} provides all neuron instances concerning the adopted vector inputs for the mentioned ANN, while each covering method must evaluate the conditions satisfied by the neuron instances of a pair of input examples. In particular, covering methods are further provided by taking instances of all specified ANN neurons and checking whether they are met. Neuron instances are represented by columns $n_{k,l}$; they can be obtained by just applying Eq.~\eqref{eq:anncalc}. For example, the output $y_{1,1}$ of $n_{1,1}$, for Ex1, as shown in Table~\ref{tab:cov}, is obtained as $y_{1,1}=0.4x_{1}+0.5x_{2}-0.2=-1.3$.
%
%

%
\begin{table}[htb]
	\caption{ANN instantiated examples and covering methods.}
	\centering
	\begin{tabular}{|l|l|l|l|l|l|l|l|}
		\hline
		\textbf{Ex} & \textbf{Input} & $n_{1,1}$ & $n_{2,1}$ & $n_{3,1}$ & $n_{1,2}$ & $n_{2,2}$ & $n_{1,3}$ \\ \hline
		Ex1 & (1, -3) & -1.3 & -1.80 & -0.50 & -0.79 & -1.370 & -1.417 \\ \hline
		Ex2 & (1, -1) & -0.3 & -0.40 & 0.10 & 0.51 & 0.090 & 0.353 \\ \hline
		Ex3 & (1, -1.2) & -0.4 & -0.54 & 0.04 & 0.38 & -0.056 & 0.176\\ \hline
		Ex4 & (1, -7) & -3.3 & -4.60 & -1.70 & -3.39 & -4.290 & -4.957 \\ \hline
	\end{tabular}
	\label{tab:cov}
\end{table}

In Table~\ref{tab:cov}, if Ex1 and Ex2 are considered as $w_1$ and $w_2$, respectively, the pair $\alpha=\{n_{3,1},n_{k,2}\}$, with $k=\{1,2\}$, is covered by SS-Cover. Here, only neuron \textit{$n_{3,1}$} has its signal changed, in layer $1$, which means that ${sc}(n_{3,1}, w_{1}, w_{2})$ and $\neg{sc}(n_{k,1}, w_{1}, w_{2}) \forall k\neq 3$ are \texttt{true}. Since only one neuron has its signal changed, in layer $1$, any neuron in layer $2$ with its signal changed will make a pair covered by SS-Cover. In this case, as the first neuron of layer $2$ ($n_{1,2}$) has its value changed, then the pair $\alpha=\{n_{3,1},n_{1,2}\}$ is SS-Covered. Ex2 and Ex3 are examples of DS-Cover. In Table~\ref{tab:cov}, there exist no signal change in layer $1$, which means that, for some metric $h$ (\eg{}, euclidean distance), the two instances of layer $1$ have a distance change and it makes $\mathrm{dc}(h,1, w_{1}, w_{2})$ equal to \texttt{true}. The distance change in layer $1$ leads to a signal change of neuron $n_{2,2}$, which makes the pair of neurons $n_{k,1}$ and $n_{2,2}$ covered by DS-Cover $\forall k\leq N_{1}$.

Ex2 and Ex3 also contain a pair covered by SV-Cover. Similar to SS-Cover, there exists only one neuron with a signal change in layer $2$, which leads to a value change on neuron $n_{1,3}$, which means that the neuron pair $\alpha=\{n_{2,2},n_{1,3}\}$ is SV-Covered. Finally, Ex1 and Ex 4 are examples of DV-Cover. In this case, there exists no signal change in any layer of the ANN instances. However, if there exists a metric that makes a distance change \texttt{true} in layer $1$ or $2$ and there also exists another metric that makes a value change \texttt{true} in any neuron of layer $2$ or $3$, then we have a DV-Covered pair.

The ANN properties to be checked by ESBMC generate a different literal for each covering method considered here. The neuron covered by one of them must be equal or greater than a percentage $P$ of all neurons contained in an ANN. In particular, these properties generate literals $l_{\mathrm{ss}}$, $l_{\mathrm{ds}}$, $l_{\mathrm{sv}}$, and $l_{\mathrm{dv}}$, with  the goal of representing the validity of the respective covering method  w.r.t. an ANN, according to four constraints: SS-Cover, DS-Cover, SV-Cover, and DV-Cover.

SS-Cover provides the most straightforward and clear signs of adversity, which happens when a pair of similar images causes signal changes in neuron instances of consecutive layers. The SS-Cover literal $l_{\mathrm{ss}}$ is defined as
 \begin{equation}
 	l_{\mathrm{ss}} \Leftrightarrow \left( \frac{\displaystyle \sum_{k,l}{{ss}(\alpha, w_{1}, w_{2})}}{NT} \geq P\right).
 	\label{eq:sscover}
 \end{equation}

\noindent where $NT$ is the total number of neurons in an ANN. DS-Cover complements SS-Cover and defines that no signal changes have occurred, but there may be a metric concerning all layer neuron instances that caused a signal change in the consecutive layer. The DS-Cover literal $l_{\mathrm{ds}}$ is defined as
 \begin{equation}
 l_{\mathrm{ds}} \Leftrightarrow \left( \frac{\displaystyle \sum_{k,l}{{ds}(n_{k,l+1},l, h,w_{1}, w_{2})}}{NT} \geq P \right).
 \label{eq:dscover}
 \end{equation}

\noindent SV-Cover complements DS-Cover and its coverage occurs when no signal change causes a signal change on the consecutive layer's neuron instances, while a significant value change is reached. The SV-Cover literal $ l_{\mathrm{sv}} $ is defined as 
 \begin{equation}
 l_{\mathrm{sv}} \Leftrightarrow \left( \frac{\displaystyle \sum_{k,l}{{sv}(\alpha, g,w_{1}, w_{2})}}{NT} \geq P\right).
 \label{eq:svcover}
 \end{equation}

\noindent Lastly, DV-Cover complements SV-Cover and is the most generalist covering method. It is defined when a pair of similar images does not cause any distance changes on a given layer's neuron instances, but a signal change is reached on its consecutive layer. Nonetheless, there may exist a significant distance change concerning a metric that reaches a value change in the consecutive layer's neuron instances. The DV-Cover literal $l_{\mathrm{dv}}$ is defined as
 \begin{equation}
 l_{\mathrm{dv}} \Leftrightarrow \left( \frac{\displaystyle \sum_{k,l}{{dv}(n_{k,l+1},l, g,h,w_{1}, w_{2})}}{NT} \geq P\right).
 \label{eq:dvcover}
 \end{equation}
 
One may notice that all four covering methods represent the adversity provided by a pair of vector inputs $w_{1}$ and $w_{2}$, for all neurons. Their literals should provide the relation of all neurons contained in layers covered by a specific covering method. As described by ~\eqref{eq:sscover}--\eqref{eq:dvcover}, this relation is checked by determining a percentage $ P $ that indicates the minimum amount of covered neurons (note that the real percentage of all covered neurons is compared with $P$).
 
\subsection{Verification of Adversarial Case}
\label{ssec:adversarial}

\begin{figure}[htb]
\centering
\includegraphics[width=0.75\textwidth]{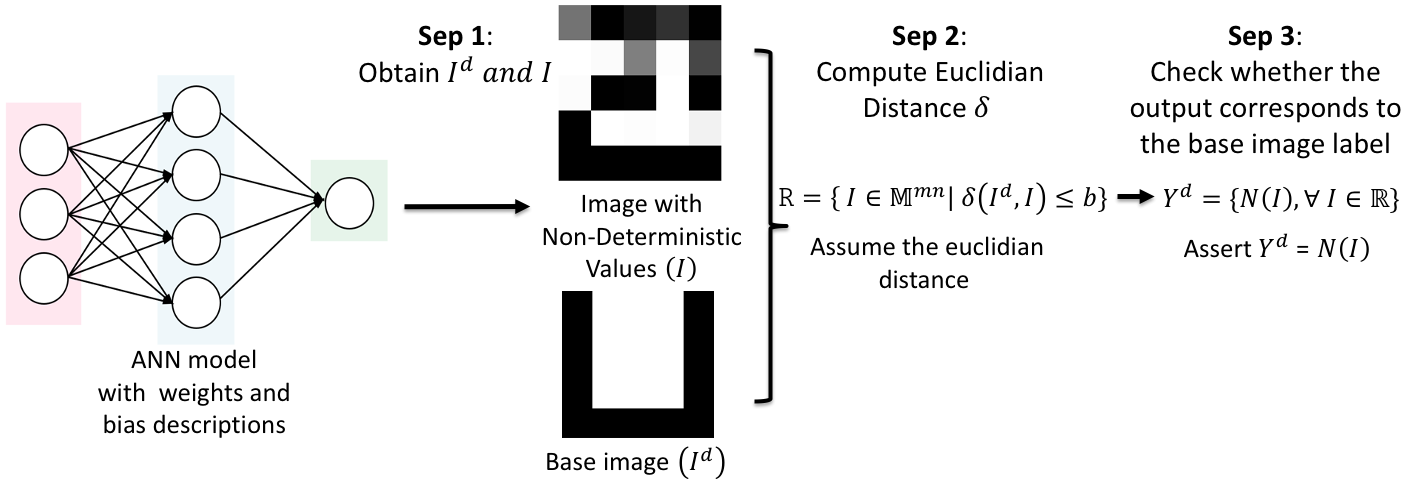}
\caption{Verification of adversarial cases.}
\label{fig:adversarial-case}
\end{figure}

Our verification algorithm can obtain an adversarial input that can lead the ANN to failures, \eg{}, misclassifying an image. Fig.~\ref{fig:adversarial-case} illustrates our verification process to obtain adversarial cases, which consists of three main steps.

\begin{figure}[htb]
	\centering
	\begin{subfigure}[b]{0.08\linewidth}
		\centering
		\includegraphics[width=\linewidth]{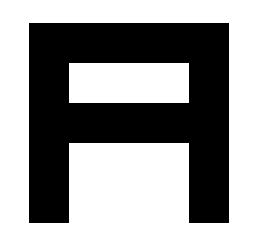}
		\label{figA:A}
	\end{subfigure}
	\begin{subfigure}[b]{0.08\linewidth}
		\centering
		\includegraphics[width=\linewidth]{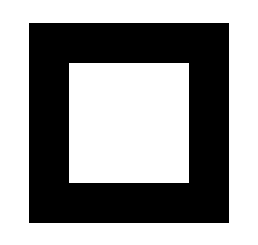}
		\label{figO:O}
	\end{subfigure}
	\caption{Vocalic ``A'' and ``O''.}\label{fig:vocalicao}
\end{figure}

Let us assume that an image input is represented by $I$, with \textit{m} and \textit{n} being its dimensions. The employed dataset is denoted as $\mathcal{D}$, while $\mathcal{M}$ is the universe of all possible images with size \textit{m}$\times$\textit{n}. In the first step of our verification process, we provide two images: an image with non-deterministic values $I$, which could represent a malicious input, and a base image $I^{d}$, from $\mathcal{D}$, to be checked. Let $\delta:\mathcal{M}^{\textit{m}\times\textit{n}}\times \mathcal{M}^{\textit{m}\times\textit{n}} \rightarrow \mathbb{R}$ be an euclidean distance operator defined as
\begin{equation}
\delta(P, Q) = \sqrt{\sum_{i=0}^{n}{(p_{i} - q_{i})^2}},
\label{eq:eucliddist}
\end{equation}

\noindent such that the images are casted into normalized vectors, and $P$ = ($p_{1}, p_{2}, p_{3}, \ldots, p_{m \times n}$) and $Q = (q_{1}, q_{2}, q_{3}, \ldots, q_{m \times n}) \in \mathbb{R}^{m \times n}$. In the second step, we use the \eqref{eq:eucliddist} to compute the euclidean distance of those two images $I$ and $I^{d}$.  Suppose that $P$ and $Q$ are $5\times 5$ images, {\it i.e.}, both vectors present a length of $25$ elements and they represent images ``A'' and ``O'', respectively, as illustrated in Fig.~\ref{fig:vocalicao}.

\noindent As an example, if we apply the euclidean distance, as described by~ \eqref{eq:eucliddist}, to the two images in Fig.~\ref{fig:vocalicao}, $\delta$ will return $2.449$. As ${I}^{d}$ belongs to $\mathcal{D}$, all sets can be represented as
\begin{equation}
\centerline{${I}^{d} \in \mathcal{D}$, ${I}$ $\in$ $\mathcal{M}^{\textit{m}\times\textit{n}}$, $\mathcal{D}^{\textit{m}\times\textit{n}}$ $\subseteq$ $\mathcal{M}^{\textit{m}\times\textit{n}}$.}
\end{equation}

\noindent The base image to be checked can be represented by an \texttt{assume}, which is formally described as
\begin{equation}
\mathcal{R} = \{ \mathcal{I} \in \mathcal{M}^{m \times n} \mid \delta (\mathcal{I}^{d},\mathcal{I}) \leq \gamma \}.
\label{eq:restriction}
\end{equation}

Thus, \eqref{eq:restriction} denotes that $\mathcal{I}^{d} \in \mathcal{M}^{m\times n}$ will be compared with a non-deterministic image $\mathcal{I}$ until the euclidean distance $\delta (\mathcal{I}^{d},\mathcal{I})$, described by  ~\eqref{eq:eucliddist}, is smaller than or equal to $\gamma$, which determines that a suitable element was found.

In the third step, after a non-deterministic image $\mathcal{I}$ is obtained, a safety property is checked, which is represented by an assertion statement within our verification engine, as
\begin{equation}
\centerline{$\mathcal{Y}^{d}$ = \{$N(\mathcal{I})$ $ ,$ $\forall$ $\mathcal{I} \in \mathcal{R}\}$.}
\label{eq:assert}
\end{equation}

\noindent In~ \eqref{eq:assert}, $\mathcal{Y}^{d}$ represents any ANN's output mapped by the input $\mathcal{I}^{d}$ on the given dataset. The function $N$ represents the ANN function $\mathbb{R}^{m} \rightarrow \mathbb{R}^{n}$, and the target property is taken as the negation of ~\eqref{eq:assert}. If the output obtained by the function $N$ and the non-deterministic image is different from the mapped output, then the property is violated and a counterexample is produced. The property described by~ \eqref{eq:assert} generates the literal \textit{$l_{\mathrm{adversarial}}$}, while a classification is obtained from the output values represented by neurons of the last layer. $ V $ is a reference value for controlling the output neurons' threshold. The desired classification is denoted by variable $ D $, which represents the neuron position in the output layer, and \textit{i} represents any other neuron position different from $D$. The literal \textit{$l_{\mathrm{adversarial}}$} represents the validity of the original image classification, according to the constraint
\begin{equation}\label{eq:imagemisclassified}
l_{\mathrm{adversarial}} \Leftrightarrow (n_{L,D} < V) \wedge \left( \bigvee_{i\neq D} (n_{L,i} \geq V)\right).
\end{equation}

\section{Experimental Evaluation}
\label{sec:exp}

\subsection{Description of the Benchmarks}
\label{ssec:benchmarks}

\begin{figure}[htb]
	\centering
	\begin{subfigure}[b]{0.08\linewidth}
		\centering
		\includegraphics[width=\linewidth]{imagema.png}
		\label{fig:A}
	\end{subfigure}
	\begin{subfigure}[b]{0.08\linewidth}
		\centering
		\includegraphics[width=\linewidth]{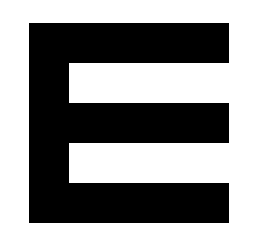}
		\label{fig:E}
	\end{subfigure}
	\begin{subfigure}[b]{0.08\linewidth}
		\centering
\includegraphics[width=\linewidth]{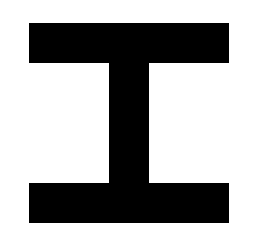}
\label{fig:I}
	\end{subfigure}
\begin{subfigure}[b]{0.08\linewidth}
		\centering
\includegraphics[width=\linewidth]{imagemo.png}
\label{fig:O}
\end{subfigure}
\begin{subfigure}[b]{0.08\linewidth}
		\centering
\includegraphics[width=\linewidth]{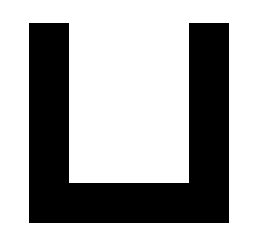}
\label{fig:U}
\end{subfigure}
	\caption{Vocalic images in benchmarks.}\label{fig:vowels}
\end{figure}

Our evaluation procedure employs a character pattern recognition benchmark~\cite{Sena20}, with the goal of performing conformance testing~\cite{KrichenT09} over our \textit{cuBLAS} and \textit{cuDNN} operational models (cf. Section~\ref{ssec:method}). In particular, each experiment employs an ANN that solves the problem of vocalic pattern recognition for images with dimensions $5\times5$, and its architecture is 25x10x4x5. The same ANN was trained with the back-propagation algorithm~\cite{bishop2006PRML} and used with a dataset composed of $100$ correct vocalics with noise and $100$ non-vocalic images. All vocalics are illustrated in Fig.~\ref{fig:vowels}.

We have conducted the experimental evaluation on a Intel(R) Xeon(R) CPU E$5$-$2620$ v$4$ @ $2$.$10$GHz with $128$ GB of RAM and Linux OS. We use CUDA v$9$.$0$, cuDNN v$5$.$0$, cuBLAS v$10$.$1$, and ESBMC v$6$.$4$.$0$.\footnote{Available at \url{http://esbmc.org/}} All presented execution times are CPU times, {\it i.e.}, only the elapsed periods spent in the allocated CPUs, which was measured with the {\it times} system call~\cite{monteiro2018esbmc}. All experimental results reported here were obtained by executing a pre-defined prompt command,\footnote{\texttt{esbmc <file.c> -I <path-to-OM> --force-malloc-success --no-div-by-zero-check --no-pointer-check --no-bounds-check --incremental-bmc --interval-analysis --fixedbv}} which is wrapped in a shell script to unwind a program iteratively.

\subsubsection{Availability of Data and Tools}
\label{ssec:AvailabilityofDataandTools}

All benchmarks, tools, and results associated with the current evaluation are available for download at \url{https://tinyurl.com/y6xvze65}.

\subsection{Objectives}
\label{ssec:desc_exp}

Our experimental evaluation has the following goals:

\begin{enumerate}

\item[EG1] \textbf{(FWL Effects)} Evaluate the peformance and correctness of our verification algorithm regarding FWL effects (cf. Section~\ref{ssec:method}) in ANN implementations.

\item[EG2] \textbf{(Covering Methods)} Evaluate the performance and correctness of our verification algorithms to validate covering methods (cf.~Section~\ref{ssec:coverage}).

\item[EG3] \textbf{(Adversarial Cases)} Compare the performance of our verification algorithm (cf. Section~\ref{ssec:adversarial}) with existing approaches, where adversarial cases are obtained by changing input images with a norm-bounded disturbance.

\end{enumerate}

\subsection{Results}
\label{ssec:results}

\subsubsection{Comparison with Existing Tools}
\label{exp:comparison}

Our verification of ANN implementations is compared to Neurify~\cite{Wang2018}, which checks for user-defined violations in neural network models. To compare our approach with it, we have applied their $3$ ANN models to the MNIST dataset~\cite{mnist}, which is used for our verification algorithm and considering fixed-point arithmetic. Moreover, the first available model, mnist$24$, has $4$ layers, in which $N_1 = 784$, $N_2 = N_3 = 24$, and $N_4 = 10$. The other two models only change the number of neurons in the second and third layers, \textit{i.e.}, $N_2 = N_3 = 50$ and $N_2 = N_3 = 512$, respectively. We have verified such models in Neurify and also in our verification algorithm, using different word lengths. On top of that, we have defined many properties regarding each neuron of each ANN's output layer. Then, we have performed a comparison regarding the evaluation of such properties under fixed-point representations.

Table~\ref{tab:comparison-mnist24} presents results for such an evaluation regarding one of the available models in Neurify~\cite{Wang2018}, \textit{i.e.}, mnist$24$, and one of its test images, \textit{i.e.}, image $1$, using our approach. Indeed, Neurify presented no violation and similar results were obtained for the other available models. \textbf{\#P} describes the proposed property to be verified regarding neuron $n_{3,m}$, \textit{i.e.}, the $m$-th neuron of the output layer; \textbf{\#N} is the  verification result provided by Neurify; \textbf{\#F\textless{}X,Y\textgreater{}} represents the verification result provided by our approach using $X$ bits for the integral part and $Y$ bits for the fractional one; and SAFE and UNSAFE denote whether an ANN implementation is safe or not, w.r.t. a specific property.

One may notice that FWL effects due to fixed-point formats affect neuron outputs until they are no longer safe, which is particularly essential for digital systems~\cite{Bessa2016}. Nonetheless, a more direct comparison could not be performed since Neurify only considers floating-point representations in its internal modeling and operations~\cite{Wang2018}. 

\begin{tcolorbox}
These results successfully answer \textbf{EG1}: our approach can identify violations concerning the word-length employed in ANN implementations.
\end{tcolorbox}

Regarding other existing tools to verify ANN implementations, {\it i.e.}, DeepConcolic~\cite{SunWRHKK18} and
DLV~\cite{huang2017safety}, we were unable to perform it directly using the same ANN in our benchmarks. The main obstacle is that such tools do not work with shallow neural networks and activation functions differently than ReLU, so as our benchmarks.

\begin{table}[htb]
\centering
\caption{Results for the proposed approach with model mnist$24$ and image $1$~\cite{Wang2018}.}
\begin{tabular}{|c|c|c|c|c|c|}
\hline
\textbf{\#P} & \textbf{\#N} & \textbf{\#F\textless{}2,2\textgreater{}} & \textbf{\#F\textless{}4,4\textgreater{}} & \textbf{\#F\textless{}8,8\textgreater{}} & \textbf{\#F\textless{}16,16\textgreater{}} \\ \hline
$n_{3,1} \leq -18.77$            & SAFE         & SAFE                                     & SAFE                                     & UNSAFE                                   & UNSAFE                                    \\ \hline
$n_{3,2} \leq -7.81$            & SAFE         & SAFE                                     & SAFE                                     & SAFE                                     & SAFE                                      \\ \hline
$n_{3,3} \leq -0.14$            & SAFE         & UNSAFE                                   & UNSAFE                                   & UNSAFE                                   & UNSAFE                                    \\ \hline
$n_{3,4} \leq -10.61$            & SAFE         & UNSAFE                                   & SAFE                                     & SAFE                                     & SAFE                                      \\ \hline
$n_{3,5} \leq -36.32$            & SAFE         & SAFE                                     & UNSAFE                                   & SAFE                                     & SAFE                                      \\ \hline
$n_{3,6} \leq -11.54$            & SAFE         & UNSAFE                                   & SAFE                                     & UNSAFE                                   & UNSAFE                                    \\ \hline
$n_{3,7} \leq -13.58$            & SAFE         & UNSAFE                                   & UNSAFE                                   & SAFE                                     & SAFE                                      \\ \hline
$n_{3,8} \leq -29.48$            & SAFE         & SAFE                                     & SAFE                                     & SAFE                                     & SAFE                                      \\ \hline
$n_{3,9} \leq -17.21$            & SAFE         & SAFE                                     & UNSAFE                                   & SAFE                                     & SAFE                                      \\ \hline
$n_{3,10} \leq -22.40$           & SAFE         & SAFE                                     & UNSAFE                                   & UNSAFE                                   & UNSAFE                                    \\ \hline
\end{tabular}
\label{tab:comparison-mnist24}
\end{table}


\subsubsection{Covering Methods}
\label{exp:CoveringMethods }

In experiments related to covering methods, even using invariant inference or not, ESBMC was able to verify all four methods correctly: SS-Cover, DS-Cover, SV-Cover, and DV-Cover (cf. Section~\ref{ssec:coverage}). The distance threshold parameter $d$ used for evaluating $g$ (the same mentioned in Section \ref{ssec:coverage} was employed here) and computing SV-cover and DV-Cover was set as $d=1$, while the distance operator in~\eqref{eq:eucliddist} is compared with $v=0.1$ for evaluating $h$, regarding DS-Cover and DV-Cover. The verification times for all four covering methods did not take longer than a few minutes for checking how adversarial two images are, w.r.t. ANN neurons, as described by~\eqref{eq:sscover}-- \eqref{eq:dvcover}, in Section~\ref{ssec:coverage}. A fast verification was indeed expected, regardless of invariant inference, since all covering methods have only deterministic inputs; however, the best performance is reached when invariant inference is used. 

\begin{tcolorbox}
These results successfully answer \textbf{EG2}: ESBMC achieves reasonable performance to validate all covering methods correctly. Moreover, using invariant inference leads to even better performance since the generated constraints reduce the resulting formulae to be checked by the underlying SMT solver (cf. Section~\ref{sssec:invariant}).
\end{tcolorbox}

In Table~\ref{tab:coveringcomparison}, \textit{Covering Method} is the experiment identifier and \textit{II} and \textit{NII} are the execution times with and without interval analysis, respectively, taken by ESBMC.

\begin{table}[htb]
	\centering
	\caption{Verification time and invariant inference relation.}
	\begin{tabular}{|c|c|c|}
		\hline
	\textbf{Covering Method} & \textbf{II (s)} & \textbf{NII (s)} \\ \hline
		SS-Cover & 819.715 & 1281.714\\ \hline 
		DS-Cover & 868.570 & 1221.256\\ \hline
		SV-Cover & 934.932 & 1709.100 \\ \hline
		DV-Cover & 978.881 & 1804.452 \\ \hline
	\end{tabular}
	\label{tab:coveringcomparison}
\end{table}

\begin{figure}[htb]
	\centering
	\begin{subfigure}[b]{0.1\linewidth}
		\centering
		\includegraphics[width=\linewidth]{imagemu.png}
		\label{fig:uori}
	\end{subfigure}
	\begin{subfigure}[b]{0.1\linewidth}
		\centering
		\includegraphics[width=\linewidth]{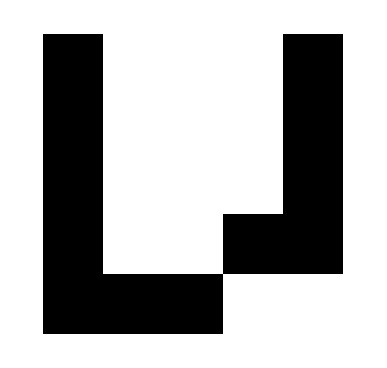}
		\label{fig:unoised}
	\end{subfigure}
	\caption{Vocalics U and U with some noise.}\label{fig:vocalicuunoised}
\end{figure}

The verified properties imply that $80$\% of all neurons on the image set must cover an ANN implementation. In particular, the image set employed here was used during the training phase; the tool's output for all benchmarks correctly returned that covered neurons were not higher than $80$\%. As a result, the dataset was unable to provide $80$\% of neuron coverage, according to any covering method. The average execution time of all four covering methods, when applied to a set of $200$ images, is around $25$ minutes, without invariant inference. The average execution time with invariant inference is around $15$ minutes, which is $60$\% faster. The activation potentials of the chosen ANN for the inputs ``U'' and noisy ``U'', illustrated in Fig.~\ref{fig:vocalicuunoised}, are presented in Table~\ref{tab:instancesU}.

\begin{table}[!h]
	\caption{ANN instances of two inputs.}
	\centering
	\begin{tabular}{|c|c|c|}
		\hline
		\textbf{Neuron} &\textbf{Image ``U''} & \textbf{Noisy ``U''} \\ \hline
		$n_{1,1}$ &-1.885322 & 4.619613 \\ \hline
		$n_{1,2}$ &8.775419 & 9.796190 \\ \hline
		$n_{1,3}$ &2.959348 & 5.743809 \\ \hline
		$n_{1,4}$ &10.424796 & 4.046428 \\ \hline
		$n_{1,5}$ &8.172012 & 14.466885 \\ \hline
		$n_{2,1}$ &-3.863095 & -9.308636 \\ \hline
		$n_{2,2}$ &5.328067 & 5.263461 \\ \hline
	\end{tabular}
	\begin{tabular}{|c|c|c|}
	\hline
	\textbf{Neuron} &\textbf{Image ``U''} & \textbf{Noisy ``U''} \\ \hline
		$n_{2,3}$ &-3.770385 & -5.705760 \\ \hline
$n_{2,4}$ &0.574238 & -2.029373 \\ \hline
$n_{3,1}$ &-6.707186 & -7.149290 \\ \hline
$n_{3,2}$ &-15.815082 & -17.246468 \\ \hline
$n_{3,3}$ &-10.060704 & -13.074245 \\ \hline
$n_{3,4}$ &-9.688183 & -4.868999 \\ \hline
$n_{3,5}$ &-0.555885 & 3.355738 \\ \hline
\end{tabular}
	\label{tab:instancesU}
\end{table}

According to the SS-Cover equation described in Section~\ref{sec:verification} and the literal described by ~\eqref{eq:sscover}, we have two pairs of the neuron ($\alpha_1=\{n_{1,1}, n_{2,4}\}$ and $\alpha_2=\{n_{2,4}, n_{3,5}\}$) SS-Covered by the two images. Only $3$ of $14$ neurons are SS-Covered; that is, only $21$\% of the neurons are covered by this method. The literal described by ~\eqref{eq:sscover} specifies that the neuron coverage must be greater than a percentage $ P $, fixed to $80$\%. It means that, for these two chosen inputs, the property fails. 

As the $200$ images of the training set do not achieve the expected coverage, as indicated by ESBMC, we have added noise to the original dataset and then rerun our validation procedure to compare the coverage results. We have also evaluated the capability of our technique to provide input examples able to increase neuron coverage w.r.t a covering method and an input vector. The verifier mutates the input vector until it finds a similar image that reaches a specified neuron coverage. We have performed our experiments with the ANN shown in Fig.~\ref{fig:motivating-example}, where the four covering methods were correctly verified. Here, our property specifies that neuron coverage must reach at least $50$\%, regarding the specified input vector and range. If the property is not satisfiable, then a counterexample is provided with a similar input vector inside the range that reaches more than $50$\% of neuron coverage. In Table~\ref{tab:coveringcomparison2},\textit{Covering Method} is the experiment identifier, \textit{II} and \textit{NII} are the execution times with and without interval analysis, respectively, taken by ESBMC. 

Fig.~\ref{fig:motivating-example} indicates that there is no any input vector pair able to increase the coverage w.r.t SS-Cover, DS-Cover, and DV-Cover. It is easy to see that the SV-Cover condition (at least $50$\%) is violated by the inputs $x=2$ and $y=2$. Furthermore, violating inputs can be obtained from counterexamples, \eg{}, using ESBMC, the inputs $x=4$ and $y = 1.3125$ are obtained. It allows us to conclude that inputs similar to $x=4$ and $y = 1.3125$ are probably able to generate a violation and an adversarial case.

\begin{tcolorbox}
Regarding \textbf{EG2}, we have also evaluated our approach for generating counterexamples to increase neuron coverage concerning the covering methods. This technique correctly verified neuron coverage properties and provided counterexamples in a reasonable time.
\end{tcolorbox}

\begin{table}[htb]
	\centering
	\caption{Verification time and verification result.}
	\begin{tabular}{|c|c|c|c|}
		\hline
		\textbf{Covering Method} & \textbf{II (s)}
		&\textbf{NII (s)} &\textbf{Verification Result}\\ \hline
		SS-Cover & 0.667 & 0.710& SAFE\\ \hline
		DS-Cover & 25.995 & 29.438& SAFE\\ \hline
		SV-Cover & 0.614 & 0.640& UNSAFE \\ \hline
		DV-Cover & 27.469 & 28.700& SAFE \\ \hline
	\end{tabular}
	\label{tab:coveringcomparison2}
\end{table}

\subsubsection{Adversarial Cases}
\label{exp:AdversarialCases}

Regarding adversarial cases, we verified $21$ benchmarks, where the average verification time was approx. $47$ hours. In this set of experiments, when enabled the \texttt{fixedbv} option in ESBMC, which forces the model checker to use a fixed-point representation of <$32$, $32$>. Table~\ref{tab:time} shows the verification time and parameter values of our experiments. Here, $B$ represents \textit{Benchmark}, which is the experiment identifier, \textit{I} represents the desired image, whose classification is verified, $\gamma$ represents a limit of proximity for each benchmark, and \textit{II} and \textit{NII} are the execution times with and without interval analysis, respectively, taken by ESBMC.

\begin{table}[htb]
	\centering
	\caption{Verification time and proximity parameter relation.}
	\begin{tabular}[t]{|c|c|c|c|c|c|}
		\hline
		\textbf{\#} & \textbf{B} &\textbf{I} & \textbf{$\gamma$} & \textbf{II (m)} & \textbf{NII (m)} \\ \hline
		1 & iae03 &  A & 0.3 & 4649 & 4651\\ \hline 
		2 & iae05 &  A & 0.5 & 4647 & 4644\\ \hline
		3 & iae10 &  A & 1.0 & 5430 & 5433\\ \hline
		4 & iae15 &  A & 1.5 & 5431 & 5426 \\ \hline
		5 & iao03 & A & 0.3 & TO &TO \\ \hline 
		6 & iao05 & A & 0.5 & TO &TO\\ \hline 
		7 & iao10 & A & 1.0 & 7597 & 7588\\ \hline 
		8 & iao15 & A & 1.5 & 7563 & 7599\\ \hline 		
		9 & ieo05 & E & 0.5 & 4951 & 4952\\ \hline 
		10 & ieo07 & E & 0.7 & 4953 &4953 \\ \hline 
		11 & ieo15 & E & 1.5 & 2283 & 2286\\ \hline 
	\end{tabular}
		\begin{tabular}[t]{|c|c|c|c|c|c|}
		\hline
		\textbf{\#} & \textbf{B} &\textbf{I} & \textbf{$\gamma$} & \textbf{II (m)} & \textbf{NII (h)} \\ \hline
		12 & ieo30 & E & 2.5 & 1819 &1818 \\ \hline
13 & iou05 & O & 0.5 & 255 & 254\\ \hline 
14 & iou07 & O & 0.7 & 255 & 255\\ \hline 
15 & iou15 & O & 1.5 & 567 & TO\\ \hline 
16 & iou30 & O & 2.5 & 954 & 953\\ \hline 
17 & iuo03 & U & 0.3 & 404 & 404\\ \hline 
18 & iuo05 & U & 0.5 & TO & 404\\ \hline 
19 & iuo10 & U & 1.0 & 745 & 743\\ \hline 
20 & iuo15 & U & 1.5 & 743 & 745\\ \hline 
21 & ieu30 & E & 2.5 & 724 & 724\\ \hline  
	\end{tabular}
	\label{tab:time}
\end{table}

The interval analysis approach has not been efficient concerning verification times since we verify the actual implementation of ANNs in CUDA, using the logics QF\_AUFBV from the SMT standard~\cite{Barrett10c}.  Although our benchmarks contain $3$ layers, ESBMC took longer than other tools, which have previously reported experimental results with larger ANNs~\cite{huang2017safety}. In particular, deep learning verification (DLV)~\cite{huang2017safety} has obtained adversarial cases of ANNs with $12$ layers, ranging from a few seconds to $20$ minutes. To check refinement by layer, DLV uses the theory of linear real arithmetic with existential and universal quantifiers. For verification within a layer (0-variation), DLV uses the same theory without universal quantification.

In prior work, Cordeiro et al.~\cite{CordeiroFM12} reported that, although verification conditions are solved faster using the theory of linear real arithmetic, since results are independent of the actual binary representation, the theory of bit-vector allows the encoding of bit-level operators more accurately, which is inherent to the implementation of ANNs. 

\begin{tcolorbox}
These results partially answer \textbf{EG3}: ESBMC can produce adversarial cases, as confirmed by graphical inspection in MATLAB, but with higher verification time, compared with DLV. Nonetheless, that is still acceptable, since finding a combination of pixels to misclassify an object and then produce an adversarial case, together with the bit-accurate precision of our verification model, is a time-consuming and challenging task.
\end{tcolorbox}

\begin{figure}[!h]
	\centering
	\captionsetup[subfigure]{labelformat=empty}
	\begin{subfigure}[t]{0.19\linewidth}
		\centering
		\begin{subfigure}[t]{0.4\linewidth}
			\centering
			\includegraphics[height=0.4in]{imagema.png}
		\end{subfigure}%
		~
		\begin{subfigure}[t]{0.4\linewidth}
			\centering
			\includegraphics[height=0.4in]{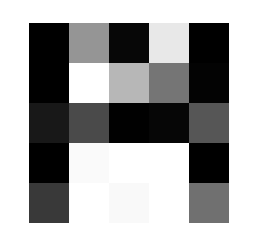}
		\end{subfigure}
		\caption{(a) "A" misclassified as "E" for $\gamma$=0.3.}
	\end{subfigure}
	\begin{subfigure}[t]{0.19\linewidth}
		\centering
		\begin{subfigure}[t]{0.4\linewidth}
			\centering
			\includegraphics[height=0.4in]{imagema.png}
		\end{subfigure}%
		~
		\begin{subfigure}[t]{0.4\linewidth}
			\centering
			\includegraphics[height=0.4in]{2iae03.png}
		\end{subfigure}
		\caption{(b) "A" misclassified as "E" for $\gamma$=0.5.}
	\end{subfigure}
	\begin{subfigure}[t]{0.19\linewidth}
		\centering
		\begin{subfigure}[t]{0.4\linewidth}
			\centering
			\includegraphics[height=0.4in]{imagema.png}
		\end{subfigure}%
		~
		\begin{subfigure}[t]{0.4\linewidth}
			\centering
			\includegraphics[height=0.4in]{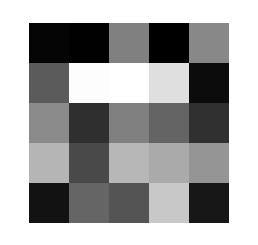}
		\end{subfigure}
		\caption{(c) "A" misclassified as "E" for $\gamma$=1.0.}
	\end{subfigure}
	\begin{subfigure}[t]{0.19\linewidth}
		\centering
		\begin{subfigure}[t]{0.4\linewidth}
			\centering
			\includegraphics[height=0.4in]{imagema.png}
		\end{subfigure}%
		~
		\begin{subfigure}[t]{0.4\linewidth}
			\centering
			\includegraphics[height=0.4in]{2iae10.png}
		\end{subfigure}
		\caption{(d) "A" misclassified as "E" for $\gamma$=1.5.}
	\end{subfigure}
	\begin{subfigure}[t]{0.19\linewidth}
		\centering
		\begin{subfigure}[t]{0.4\linewidth}
			\centering
			\includegraphics[height=0.4in]{imageme.png}
		\end{subfigure}%
		~
		\begin{subfigure}[t]{0.4\linewidth}
			\centering
			\includegraphics[height=0.4in]{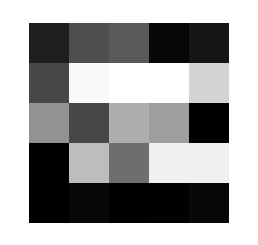}
		\end{subfigure}
		\caption{(e) "E" misclassified as "O" with $\gamma$=0.5.}
	\end{subfigure}
	\begin{subfigure}[t]{0.19\linewidth}
		\centering
		\begin{subfigure}[t]{0.4\linewidth}
			\centering
			\includegraphics[height=0.4in]{imageme.png}
		\end{subfigure}%
		~
		\begin{subfigure}[t]{0.4\linewidth}
			\centering
			\includegraphics[height=0.4in]{2ieo05.png}
		\end{subfigure}
		\caption{(f) "E" misclassified as "O" with $\gamma$=0.7.}
	\end{subfigure}
	\begin{subfigure}[t]{0.19\linewidth}
		\centering
		\begin{subfigure}[t]{0.4\linewidth}
			\centering
			\includegraphics[height=0.4in]{imageme.png}
		\end{subfigure}%
		~
		\begin{subfigure}[t]{0.4\linewidth}
			\centering
			\includegraphics[height=0.4in]{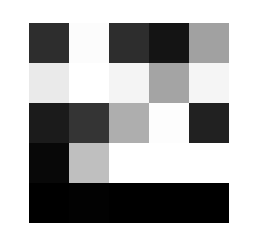}
		\end{subfigure}
		\caption{(g) "E" misclassified as "O" with $\gamma$=1.5.}
	\end{subfigure}
	\begin{subfigure}[t]{0.19\linewidth}
		\centering
		\begin{subfigure}[t]{0.4\linewidth}
			\centering
			\includegraphics[height=0.4in]{imageme.png}
		\end{subfigure}%
		~
		\begin{subfigure}[t]{0.4\linewidth}
			\centering
			\includegraphics[height=0.4in]{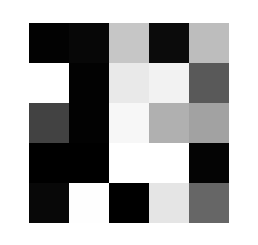}
		\end{subfigure}
		\caption{(h) "E" misclassified as "O" with $\gamma$=3.0.}
	\end{subfigure}
	\begin{subfigure}[t]{0.19\linewidth}
	\centering
	\begin{subfigure}[t]{0.4\linewidth}
		\centering
		\includegraphics[height=0.4in]{imagema.png}
	\end{subfigure}%
	~
	\begin{subfigure}[t]{0.4\linewidth}
		\centering
		\includegraphics[height=0.4in]{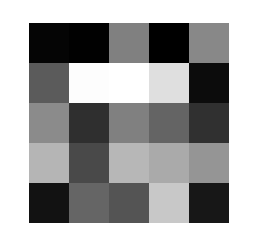}
	\end{subfigure}
	\caption{(i) "A" misclassified as "O" with $\gamma$=1.0.}
	\end{subfigure}
\begin{subfigure}[t]{0.19\linewidth}
	\centering
	\begin{subfigure}[t]{0.4\linewidth}
		\centering
		\includegraphics[height=0.4in]{imagema.png}
	\end{subfigure}%
	~
	\begin{subfigure}[t]{0.4\linewidth}
		\centering
		\includegraphics[height=0.4in]{2iao10.png}
	\end{subfigure}
	\caption{(j) "A" misclassified as "O" with $\gamma$=1.5.}
\end{subfigure}
\begin{subfigure}[t]{0.19\linewidth}
	\centering
	\begin{subfigure}[t]{0.4\linewidth}
		\centering
		\includegraphics[height=0.4in]{imagemu.png}
	\end{subfigure}%
	~
	\begin{subfigure}[t]{0.4\linewidth}
		\centering
		\includegraphics[height=0.4in]{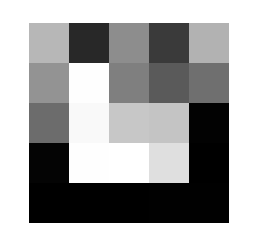}
	\end{subfigure}
	\caption{(k) "U" misclassified as "O" with $\gamma$=1.0.}
\end{subfigure}
\begin{subfigure}[t]{0.19\linewidth}
	\centering
	\begin{subfigure}[t]{0.4\linewidth}
		\centering
		\includegraphics[height=0.4in]{imagemu.png}
	\end{subfigure}%
	~
	\begin{subfigure}[t]{0.4\linewidth}
		\centering
		\includegraphics[height=0.4in]{2iuo10.png}
	\end{subfigure}
	\caption{(l) "U" misclassified as "O" with $\gamma$=1.5.}
\end{subfigure}
\begin{subfigure}[t]{0.19\linewidth}
	\centering
	\begin{subfigure}[t]{0.4\linewidth}
		\centering
		\includegraphics[height=0.4in]{imagemo.png}
	\end{subfigure}%
	~
	\begin{subfigure}[t]{0.4\linewidth}
		\centering
		\includegraphics[height=0.4in]{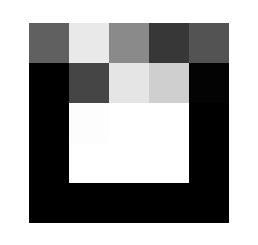}
	\end{subfigure}
	\caption{(m) "O" misclassified as "U" with $\gamma$=0.5.}
\end{subfigure}
\begin{subfigure}[t]{0.19\linewidth}
	\centering
	\begin{subfigure}[t]{0.4\linewidth}
		\centering
		\includegraphics[height=0.4in]{imagemo.png}
	\end{subfigure}%
	~
	\begin{subfigure}[t]{0.4\linewidth}
		\centering
		\includegraphics[height=0.4in]{2iou05.png}
	\end{subfigure}
	\caption{(n) "O" misclassified as "U" with $\gamma$=0.7.}
\end{subfigure}
\begin{subfigure}[t]{0.19\linewidth}
	\begin{subfigure}[t]{0.4\linewidth}
		\centering
		\includegraphics[height=0.4in]{imagema.png}
	\end{subfigure}%
	~
	\begin{subfigure}[t]{0.4\linewidth}
		\centering
		\includegraphics[height=0.4in]{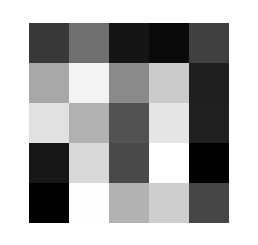}
	\end{subfigure}
	\caption{(o) "A" misclassified as "O" with $\gamma$=1.0.}
\end{subfigure}
%
%
	\caption{Generated adversarial cases.}
	\label{res:adver}
\end{figure}

In future work, we will tackle verification-time reduction to evaluate our approach with more benchmarks. Some of the adversarial cases produced by ESBMC are illustrated in Fig.~\ref{res:adver}.


\subsection{Threats to validity}
\label{sec:ThreatstoValidity}

Although all adversarial cases obtained from the benchmarks are real, which was confirmed by executing our validation scripts in MATLAB, our algorithm uses a lookup table to compute the activation function \textit{sigmoid}. Even with a proper resolution, a lookup table will always contain more value errors than a computed function, which could produce a false adversarial case. In our experiments, we used a lookup table with three decimal places resolution, but it may not be appropriated to other ANNs experiments. Additionally, it is probably easy to find adversarial cases for our character pattern recognition benchmark, since its training was performed by a small dataset, which contains only $200$ images. This kind of training probably generates ANNs with a low-level of safety classification. Finally, we were unable to check larger ANNs due to the dimensionality of the SMT formulae that exceeded the solvers' limits.

\section{Related work}
\label{sec:related_work}

Our ultimate goal is to formally ensure safety for applications
based on Artificial Intelligence (AI), as described by
Amodei \etal{}~\cite{amodei2016concrete}. In particular, the potential impact of intelligent systems performing tasks in
society and how safety guarantees are necessary to prevent
damages are the main problem of safety in ANNs.

Sun \etal{}~\cite{SunWRHKK18} and Huang \etal{}~\cite{huang2017safety} have exposed ANNs weaknesses regarding input noise. They described and
evaluated testing approaches based on covering methods and
image proximity~\cite{SunWRHKK18}, while tackling how adversarial cases are
obtained~\cite{huang2017safety}. In particular, our study
resembles those~\cite{huang2017safety,SunWRHKK18}, regarding creation of adversarial
cases. Here, if any property is violated, then a counterexample
is provided. In cases of safety properties, adversarial examples
will be generated via counterexamples output by ESBMC.
In contrast to Huang \etal{}~\cite{huang2017safety}, we do not focus
on adding noise to specific regions, but indeed to every image
pixel. Our approach regarding image proximity was influenced by Sun \etal{}~\cite{SunWRHKK18}, but we have used incremental BMC instead of
concolic testing. Our symbolic
verification method checks safety properties on
non-deterministic images with a certain distance from a given
base image; users can configure both image and distance.
Gopinath \etal{}~\cite{abs-1807-10439} also describe an approach
to validate ANNs, but using symbolic execution and translating an ANN
into an imperative program. By contrast, we consider the actual
implementation of an ANN in CUDA and apply incremental BMC using
off-the-shelf SMT solvers.

Gopinath \etal{}~\cite{DBLP:journals/corr/abs-1904-13215}
presented formal techniques to extract invariants from the
decision logic of ANNs, which represent pre- and
post-conditions that hold when transformations of a certain
type are applied to ANNs. The authors have proposed two
techniques. The first one is called iterative relaxation of
decision patterns and uses Reluplex as decision
procedure~\cite{katz2017reluplex}. The second one, in turn, is called
decision-tree based invariant generation, which resembles
covering methods~\cite{SunWRHKK18}. Robustness and
explainability are the core properties of this study, and applying
those properties to ANNs have shown impressive experimental
results. Explainability showed an important property to evaluate
safety in ANNs: the core idea is to obtain an explanation for
an adversarial case, by observing the pattern
activation behaviour of a subset of neurons described by a
given invariant.

Gopinath \etal{}~\cite{DBLP:journals/corr/abs-1710-00486} also
proposed an approach for automatically identifying safe
regions of inputs, concerning some labels. The core idea is to
identify safe regions concerning labelled targets. As
the notion of robustness in ANNs is a strong term for
many of them, the target robustness is the leading property. That
technique works with clustering and verification: a clustering
technique is used to split the dataset into a subset of inputs
with the same labels, while each cluster is then verified by
Reluplex~\cite{katz2017reluplex} in order to provide the safety region
concerning a target label. 

Wang \etal{}~\cite{Wang2018} propose an efficient approach for checking different safety properties of large neural networks, aiming at finding adversarial cases. Their approach is based on two main ideas. First, symbolic linear relaxation, which combines symbolic interval analysis and linear relaxation to create an efficient propagation method for providing tighter estimations. Second, directed-constraint refinement, which identifies nodes whose output is overestimated and iteratively refines their output ranges. Those techniques are implemented in a tool called Neurify that was validated against multiple ANN architectures. Furthermore, in order to scale up their verification framework, they have implemented their code using multi-threaded programming techniques. Indeed, the approach proposed here was evaluated againts Neurify and provided interesting results, with FWL effects.

In addition to ESBMC, there exist other tools able to verify
CUDA programs and each one of them uses its approach and targets
specific property violations. Nonetheless, given the current
knowledge in software verification, ESBMC is the first
verifier to obtain adversarial cases and validate coverage
methods in ANNs implemented with CUDA. For instance,
GPUVerify~\cite{BettsCDQT12} is based on synchronous and delayed
visibility semantics, which focuses on detecting data race and
barrier divergence, while reducing kernel verification
procedures for the analysis of sequential programs. GPU+KLEE
(GKLEE)~\cite{LiLSGGR12}, in turn, is a concrete and symbolic
execution tool, which considers both {\it kernels} and {\it
main} functions, while checking deadlocks, memory coalescing,
data race, warp divergence, and compilation level issues. Also,
concurrency intermediate verification language
(CIVL)~\cite{ZhengRLDS15}, a framework for static analysis and
concurrent program verification, uses abstract syntax tree and
partial-order reduction to detect user-specified assertions,
deadlocks, memory leaks, invalid pointer dereference, array
out-of-bounds, and division by zero.

Our approach implemented on top of ESBMC has some
similarities with other techniques described here, \eg{},
regarding the covering methods proposed by Sun \etal{}~\cite{SunWRHKK18}, model
checking to obtain adversarial cases proposed by Huang \etal{}\cite{huang2017safety}, and incremental verification of ANNs
implemented in CUDA by Sena \etal{}~\cite{Sena20}; however, the main contribution concerns our
requirements and how we handle, with invariant inference, the actual implementations of
ANNs, also taking into account FWL effects. Moreover, the latter results in promptly deployable ANNs, which could be integrated into a unified design framework. To run our proposed safety verification, 
only an ANNs' weights, bias descriptors, and desired input, regarding a
dataset, are required. For tools such as
DeepConcolic~\cite{SunWRHKK18} and DLV~\cite{huang2017safety},
obtaining adversarial cases or safety guarantees in customized ANNs depends on the intrisic characteristics of the models. For instance, in their implementations, they do not support the \textit{sigmoid} activation function or they do not support shallow neural networks, which is crucial for our ANN model to be verified properly. In our proposed approach, users are free to provide a desired dataset sample to be verified. Besides those requirements, it is necessary for a user to know how cuDNN~\cite{chetlur2014cudnn} deals with ANNs. Moreover, Sena \etal{}~\cite{Sena20} do not exploit invariant inference  to prune the state space exploration, which is done in our proposed approach.

\section{Conclusions}
\label{sec:conclusion}

The verification of ANNs has recently attracted lots of attention,  with notable approaches from formal~\cite{huang2017safety} and coverage-guided verification~\cite{SunWRHKK18}.  While existing verification methods work with neural network models and adversarial examples  (\eg{}, a small perturbation on a correctly-labelled input leads to a different classification), it has been pointed out in \cite{pmlr-v97-odena19a} that there exist errors in the Tensorflow graph representation of ANNs,  a lower-level implementation of them, such as numerical errors and disagreements between neural network implementations and their  quantized versions. 

We have described and evaluated two approaches for verifying actual implementations of ANNs, {\it i.e.}, when working with code-level implementations:  one to obtain adversarial cases and another to validate coverage methods, in MLP. In particular, our validation of covering methods was able to verify our dataset correctly and has shown to be effective during our experiments, leading to a speed-up of one order of magnitude. Our verification method was also able to find adversarial cases for different input images and proximity parameter values based on incremental BMC and invariant inference, although at the cost of higher verification times when compared with DLV. From $21$ experiments, the proposed approach was able to find adversarial cases in $18$ of them, while $3$ benchmarks timed out. In other approaches, obtaining adversarial cases of even bigger ANNs tends to be faster~\cite{huang2017safety}, but noise is not fully explored in code-level implementations, which can thus miss adversarial cases.

Besides, one aspect of the proposed approach fulfilled a gap in ANN verification: consideration of FWL effects. We have shown that their cumulative influence may compromise safety verification and are not even taken into account by existing tools, such as Neurify, which only considers floating-point arithmetic. For future work, we will investigate fault localization and repair techniques~\cite{AlvesCF17} to explain errors and make the ANN implementation robust against small noises present in the ANN inputs. Moreover, we will revisit the adversarial case generation in order to speed up this process.


\begin{thebibliography}{53}


\ifx \showCODEN    \undefined \def \showCODEN     #1{\unskip}     \fi
\ifx \showDOI      \undefined \def \showDOI       #1{#1}\fi
\ifx \showISBNx    \undefined \def \showISBNx     #1{\unskip}     \fi
\ifx \showISBNxiii \undefined \def \showISBNxiii  #1{\unskip}     \fi
\ifx \showISSN     \undefined \def \showISSN      #1{\unskip}     \fi
\ifx \showLCCN     \undefined \def \showLCCN      #1{\unskip}     \fi
\ifx \shownote     \undefined \def \shownote      #1{#1}          \fi
\ifx \showarticletitle \undefined \def \showarticletitle #1{#1}   \fi
\ifx \showURL      \undefined \def \showURL       {\relax}        \fi
\providecommand\bibfield[2]{#2}
\providecommand\bibinfo[2]{#2}
\providecommand\natexlab[1]{#1}
\providecommand\showeprint[2][]{arXiv:#2}

\bibitem[\protect\citeauthoryear{??}{10.}{2014}]%
        {10.5555/2935593}
 \bibinfo{year}{2014}\natexlab{}.
\newblock \bibinfo{booktitle}{\emph{Professional CUDA C Programming}
  (\bibinfo{edition}{1st} ed.)}.
\newblock \bibinfo{publisher}{Wrox Press Ltd.}, \bibinfo{address}{GBR}.
\newblock
\showISBNx{1118739329}


\bibitem[\protect\citeauthoryear{Abreu, Gadelha, Cordeiro, de~Lima~Filho, and
  da~Silva~Jr.}{Abreu et~al\mbox{.}}{2016}]%
        {AbreuGCFS16}
\bibfield{author}{\bibinfo{person}{Renato~B. Abreu}, \bibinfo{person}{Mikhail
  Y.~R. Gadelha}, \bibinfo{person}{Lucas~C. Cordeiro},
  \bibinfo{person}{Eddie~Batista de Lima~Filho}, {and}
  \bibinfo{person}{Waldir~Sabino da Silva~Jr.}}
  \bibinfo{year}{2016}\natexlab{}.
\newblock \showarticletitle{Bounded model checking for fixed-point digital
  filters}.
\newblock \bibinfo{journal}{\emph{J. Braz. Comput. Soc.}} \bibinfo{volume}{22},
  \bibinfo{number}{1} (\bibinfo{year}{2016}), \bibinfo{pages}{1:1--1:20}.
\newblock


\bibitem[\protect\citeauthoryear{Alves, Cordeiro, and Filho}{Alves
  et~al\mbox{.}}{2017}]%
        {AlvesCF17}
\bibfield{author}{\bibinfo{person}{Erickson Alves}, \bibinfo{person}{Lucas
  Cordeiro}, {and} \bibinfo{person}{Eddie~Lima Filho}.}
  \bibinfo{year}{2017}\natexlab{}.
\newblock \showarticletitle{A method to localize faults in concurrent {C}
  programs}.
\newblock \bibinfo{journal}{\emph{Journal of Systems and Software}}
  \bibinfo{volume}{132} (\bibinfo{year}{2017}), \bibinfo{pages}{336--352}.
\newblock


\bibitem[\protect\citeauthoryear{Amato, L{\'o}pez, Pe{\~n}a-M{\'e}ndez,
  Va{\v{n}}hara, Hampl, and Havel}{Amato et~al\mbox{.}}{2013}]%
        {amato2013artificial}
\bibfield{author}{\bibinfo{person}{Filippo Amato}, \bibinfo{person}{Alberto
  L{\'o}pez}, \bibinfo{person}{Eladia~Mar{\'\i}a Pe{\~n}a-M{\'e}ndez},
  \bibinfo{person}{Petr Va{\v{n}}hara}, \bibinfo{person}{Ale{\v{s}} Hampl},
  {and} \bibinfo{person}{Josef Havel}.} \bibinfo{year}{2013}\natexlab{}.
\newblock \showarticletitle{Artificial neural networks in medical diagnosis}.
\newblock \bibinfo{journal}{\emph{Journal of Applied Biomedicine}}
  \bibinfo{volume}{11}, \bibinfo{number}{2} (\bibinfo{year}{2013}),
  \bibinfo{pages}{47 -- 58}.
\newblock


\bibitem[\protect\citeauthoryear{Amodei, Olah, Steinhardt, Christiano,
  Schulman, and Man\'e}{Amodei et~al\mbox{.}}{2016}]%
        {amodei2016concrete}
\bibfield{author}{\bibinfo{person}{Dario Amodei}, \bibinfo{person}{Chris Olah},
  \bibinfo{person}{Jacob Steinhardt}, \bibinfo{person}{Paul Christiano},
  \bibinfo{person}{John Schulman}, {and} \bibinfo{person}{Dan Man\'e}.}
  \bibinfo{year}{2016}\natexlab{}.
\newblock \showarticletitle{Concrete Problems in AI Safety}.
\newblock  (\bibinfo{date}{06} \bibinfo{year}{2016}).
\newblock


\bibitem[\protect\citeauthoryear{Barrett, Stump, and Tinelli}{Barrett
  et~al\mbox{.}}{2010}]%
        {Barrett10c}
\bibfield{author}{\bibinfo{person}{Clark Barrett}, \bibinfo{person}{Aaron
  Stump}, {and} \bibinfo{person}{Cesare Tinelli}.}
  \bibinfo{year}{2010}\natexlab{}.
\newblock \bibinfo{booktitle}{\emph{{T}he {SMT}-{LIB} {S}tandard: {V}ersion
  2.0}}.
\newblock \bibinfo{type}{{T}echnical {R}eport}.
\newblock


\bibitem[\protect\citeauthoryear{Betts, Chong, Donaldson, Qadeer, and
  Thomson}{Betts et~al\mbox{.}}{2012}]%
        {BettsCDQT12}
\bibfield{author}{\bibinfo{person}{Adam Betts}, \bibinfo{person}{Nathan Chong},
  \bibinfo{person}{Alastair Donaldson}, \bibinfo{person}{Shaz Qadeer}, {and}
  \bibinfo{person}{Paul Thomson}.} \bibinfo{year}{2012}\natexlab{}.
\newblock \showarticletitle{{GPUVerify:} a verifier for {GPU} kernels}. In
  \bibinfo{booktitle}{\emph{27th Annual Conference on Object-Oriented
  Programming, Systems, Languages, and Applications}}.
  \bibinfo{pages}{113--132}.
\newblock


\bibitem[\protect\citeauthoryear{Biere, Cimatti, Clarke, Fujita, and Zhu}{Biere
  et~al\mbox{.}}{1999}]%
        {BiereCCFZ99}
\bibfield{author}{\bibinfo{person}{Armin Biere}, \bibinfo{person}{Alessandro
  Cimatti}, \bibinfo{person}{Edmund Clarke}, \bibinfo{person}{Masahiro Fujita},
  {and} \bibinfo{person}{Yunshan Zhu}.} \bibinfo{year}{1999}\natexlab{}.
\newblock \showarticletitle{Symbolic Model Checking Using {SAT} Procedures
  instead of {BDDs}}. In \bibinfo{booktitle}{\emph{36th annual Design
  Automation Conference}}. \bibinfo{pages}{317--320}.
\newblock


\bibitem[\protect\citeauthoryear{Bishop}{Bishop}{2006}]%
        {bishop2006PRML}
\bibfield{author}{\bibinfo{person}{Christopher Bishop}.}
  \bibinfo{year}{2006}\natexlab{}.
\newblock \bibinfo{booktitle}{\emph{Pattern Recognition and Machine Learning}}.
\newblock \bibinfo{publisher}{Springer}.
\newblock


\bibitem[\protect\citeauthoryear{Bojarski, Testa, Dworakowski, Firner, Flepp,
  Goyal, Jackel, Monfort, Muller, Zhang, Zhang, Zhao, and Zieba}{Bojarski
  et~al\mbox{.}}{2016}]%
        {bojarski2016end}
\bibfield{author}{\bibinfo{person}{Mariusz Bojarski}, \bibinfo{person}{Davide
  Testa}, \bibinfo{person}{Daniel Dworakowski}, \bibinfo{person}{Bernhard
  Firner}, \bibinfo{person}{Beat Flepp}, \bibinfo{person}{Prasoon Goyal},
  \bibinfo{person}{Larry Jackel}, \bibinfo{person}{Mathew Monfort},
  \bibinfo{person}{Urs Muller}, \bibinfo{person}{Jiakai Zhang},
  \bibinfo{person}{Xin Zhang}, \bibinfo{person}{Jake Zhao}, {and}
  \bibinfo{person}{Karol Zieba}.} \bibinfo{year}{2016}\natexlab{}.
\newblock \showarticletitle{End to End Learning for Self-Driving Cars}.
\newblock  (\bibinfo{date}{04} \bibinfo{year}{2016}).
\newblock


\bibitem[\protect\citeauthoryear{Chaves, Bessa, Ismail, dos Santos~Frutuoso,
  Cordeiro, and de~Lima~Filho}{Chaves et~al\mbox{.}}{2018}]%
        {ChavesBIFCF18}
\bibfield{author}{\bibinfo{person}{Lennon~C. Chaves}, \bibinfo{person}{Iury
  Bessa}, \bibinfo{person}{Hussama Ismail}, \bibinfo{person}{Adriano~Bruno dos
  Santos~Frutuoso}, \bibinfo{person}{Lucas~C. Cordeiro}, {and}
  \bibinfo{person}{Eddie~Batista de Lima~Filho}.}
  \bibinfo{year}{2018}\natexlab{}.
\newblock \showarticletitle{DSVerifier-Aided Verification Applied to Attitude
  Control Software in Unmanned Aerial Vehicles}.
\newblock \bibinfo{journal}{\emph{{IEEE} Trans. Reliab.}} \bibinfo{volume}{67},
  \bibinfo{number}{4} (\bibinfo{year}{2018}), \bibinfo{pages}{1420--1441}.
\newblock


\bibitem[\protect\citeauthoryear{Chaves, Ismail, Bessa, Cordeiro, and
  de~Lima~Filho}{Chaves et~al\mbox{.}}{2019}]%
        {Bessa2016}
\bibfield{author}{\bibinfo{person}{Lennon~C. Chaves},
  \bibinfo{person}{Hussama~I. Ismail}, \bibinfo{person}{Iury~V. Bessa},
  \bibinfo{person}{Lucas~C. Cordeiro}, {and} \bibinfo{person}{Eddie~B. de
  Lima~Filho}.} \bibinfo{year}{2019}\natexlab{}.
\newblock \showarticletitle{Verifying fragility in digital systems with
  uncertainties using DSVerifier v2.0}.
\newblock \bibinfo{journal}{\emph{J Syst Softw}} \bibinfo{volume}{153},
  \bibinfo{number}{2019} (\bibinfo{year}{2019}), \bibinfo{pages}{22--43}.
\newblock


\bibitem[\protect\citeauthoryear{Chetlur, Woolley, Vandermersch, Cohen, Tran,
  Catanzaro, and Shelhamer}{Chetlur et~al\mbox{.}}{2014}]%
        {chetlur2014cudnn}
\bibfield{author}{\bibinfo{person}{Sharan Chetlur}, \bibinfo{person}{Cliff
  Woolley}, \bibinfo{person}{Philippe Vandermersch}, \bibinfo{person}{Jonathan
  Cohen}, \bibinfo{person}{John Tran}, \bibinfo{person}{Bryan Catanzaro}, {and}
  \bibinfo{person}{Evan Shelhamer}.} \bibinfo{year}{2014}\natexlab{}.
\newblock \showarticletitle{cuDNN: Efficient Primitives for Deep Learning}.
\newblock  (\bibinfo{date}{10} \bibinfo{year}{2014}).
\newblock


\bibitem[\protect\citeauthoryear{Cordeiro, Fischer, and
  Marques{-}Silva}{Cordeiro et~al\mbox{.}}{2012}]%
        {CordeiroFM12}
\bibfield{author}{\bibinfo{person}{Lucas Cordeiro}, \bibinfo{person}{Bernd
  Fischer}, {and} \bibinfo{person}{Jo{\~{a}}o Marques{-}Silva}.}
  \bibinfo{year}{2012}\natexlab{}.
\newblock \showarticletitle{SMT-Based Bounded Model Checking for Embedded
  {ANSI-C} Software}.
\newblock \bibinfo{journal}{\emph{{IEEE} Trans. Software Eng.}}
  \bibinfo{volume}{38}, \bibinfo{number}{4} (\bibinfo{year}{2012}),
  \bibinfo{pages}{957--974}.
\newblock


\bibitem[\protect\citeauthoryear{Eykholt, Evtimov, Fernandes, Li, Rahmati,
  Xiao, Prakash, Kohno, and Song}{Eykholt et~al\mbox{.}}{2018}]%
        {DBLP:conf/cvpr/EykholtEF0RXPKS18}
\bibfield{author}{\bibinfo{person}{Kevin Eykholt}, \bibinfo{person}{Ivan
  Evtimov}, \bibinfo{person}{Earlence Fernandes}, \bibinfo{person}{Bo Li},
  \bibinfo{person}{Amir Rahmati}, \bibinfo{person}{Chaowei Xiao},
  \bibinfo{person}{Atul Prakash}, \bibinfo{person}{Tadayoshi Kohno}, {and}
  \bibinfo{person}{Dawn Song}.} \bibinfo{year}{2018}\natexlab{}.
\newblock \showarticletitle{Robust Physical-World Attacks on Deep Learning
  Visual Classification}. In \bibinfo{booktitle}{\emph{Conference on Computer
  Vision and Pattern Recognition}}. \bibinfo{pages}{1625--1634}.
\newblock


\bibitem[\protect\citeauthoryear{Gadelha, Ismail, and Cordeiro}{Gadelha
  et~al\mbox{.}}{2017b}]%
        {esbmc-sttt-2017}
\bibfield{author}{\bibinfo{person}{Mikhail Gadelha}, \bibinfo{person}{Hussama
  Ismail}, {and} \bibinfo{person}{Lucas Cordeiro}.}
  \bibinfo{year}{2017}\natexlab{b}.
\newblock \showarticletitle{Handling loops in bounded model checking of {C}
  programs via {k-induction}}.
\newblock \bibinfo{journal}{\emph{Software Tools for Technology Transfer}}
  \bibinfo{volume}{19}, \bibinfo{number}{1} (\bibinfo{year}{2017}),
  \bibinfo{pages}{97--114}.
\newblock


\bibitem[\protect\citeauthoryear{Gadelha, Menezes, Monteiro, Cordeiro, and
  Nicole}{Gadelha et~al\mbox{.}}{[n.d.]}]%
        {esbmc-fase-2020}
\bibfield{author}{\bibinfo{person}{Mikhail Gadelha}, \bibinfo{person}{Rafael
  Menezes}, \bibinfo{person}{Felipe Monteiro}, \bibinfo{person}{Lucas
  Cordeiro}, {and} \bibinfo{person}{Denis Nicole}.}
  \bibinfo{year}{[n.d.]}\natexlab{}.
\newblock \showarticletitle{ESBMC: Scalable and Precise Test-Case Generation
  based on the Floating-Point Theory}. In \bibinfo{booktitle}{\emph{23rd
  International Conference on Fundamental Approaches to Software Engineering
  (FASE)}}.
\newblock


\bibitem[\protect\citeauthoryear{Gadelha, Monteiro, Cordeiro, and
  Nicole}{Gadelha et~al\mbox{.}}{2019}]%
        {GadelhaMCN19}
\bibfield{author}{\bibinfo{person}{Mikhail Gadelha}, \bibinfo{person}{Felipe
  Monteiro}, \bibinfo{person}{Lucas Cordeiro}, {and} \bibinfo{person}{Denis
  Nicole}.} \bibinfo{year}{2019}\natexlab{}.
\newblock \showarticletitle{{ESBMC} v6.0: Verifying {C} Programs Using
  {k-Induction} and Invariant Inference - (Competition Contribution)}. In
  \bibinfo{booktitle}{\emph{28th International Conference on Tools and
  Algorithms for the Construction and Analysis of Systems}}.
  \bibinfo{pages}{209--213}.
\newblock


\bibitem[\protect\citeauthoryear{Gadelha, Cordeiro, and Nicole}{Gadelha
  et~al\mbox{.}}{2017a}]%
        {GadelhaCN17}
\bibfield{author}{\bibinfo{person}{Mikhail Y.~R. Gadelha},
  \bibinfo{person}{Lucas~C. Cordeiro}, {and} \bibinfo{person}{Denis~A.
  Nicole}.} \bibinfo{year}{2017}\natexlab{a}.
\newblock \showarticletitle{Encoding Floating-Point Numbers Using the {SMT}
  Theory in {ESBMC:} An Empirical Evaluation over the {SV-COMP} Benchmarks}. In
  \bibinfo{booktitle}{\emph{20th Brazilian Symposium on Formal Methods:
  Foundations and Applications (SBMF)}} \emph{(\bibinfo{series}{LNCS},
  Vol.~\bibinfo{volume}{10623})}. \bibinfo{publisher}{Springer},
  \bibinfo{pages}{91--106}.
\newblock


\bibitem[\protect\citeauthoryear{Gadelha, Cordeiro, and Nicole}{Gadelha
  et~al\mbox{.}}{2020}]%
        {abs-2004-12699}
\bibfield{author}{\bibinfo{person}{Mikhail Y.~R. Gadelha},
  \bibinfo{person}{Lucas~C. Cordeiro}, {and} \bibinfo{person}{Denis~A.
  Nicole}.} \bibinfo{year}{2020}\natexlab{}.
\newblock \showarticletitle{An Efficient Floating-Point Bit-Blasting {API} for
  Verifying {C} Programs}.
\newblock \bibinfo{journal}{\emph{CoRR}}  \bibinfo{volume}{abs/2004.12699}
  (\bibinfo{year}{2020}).
\newblock


\bibitem[\protect\citeauthoryear{Google}{Google}{2019}]%
        {tensor-flow}
\bibfield{author}{\bibinfo{person}{Google}.} \bibinfo{year}{2019}\natexlab{}.
\newblock \bibinfo{title}{Tensor Flow}.
\newblock \bibinfo{howpublished}{\url{https://www.tensorflow.org/}}.
\newblock
\newblock
\shownote{[Online; accessed August-2019].}


\bibitem[\protect\citeauthoryear{G{\"{u}}nther and Weissenbacher}{G{\"{u}}nther
  and Weissenbacher}{2014}]%
        {GuntherW14}
\bibfield{author}{\bibinfo{person}{Henning G{\"{u}}nther} {and}
  \bibinfo{person}{Georg Weissenbacher}.} \bibinfo{year}{2014}\natexlab{}.
\newblock \showarticletitle{Incremental bounded software model checking}. In
  \bibinfo{booktitle}{\emph{21st International {SPIN} Symposium on Model
  Checking of Software}}. \bibinfo{pages}{40--47}.
\newblock


\bibitem[\protect\citeauthoryear{Hagan, Demuth, Beale, and De~Jes{\'u}s}{Hagan
  et~al\mbox{.}}{1996}]%
        {hagan1996neural}
\bibfield{author}{\bibinfo{person}{Martin Hagan}, \bibinfo{person}{Howard
  Demuth}, \bibinfo{person}{Mark Beale}, {and} \bibinfo{person}{Orlando
  De~Jes{\'u}s}.} \bibinfo{year}{1996}\natexlab{}.
\newblock \bibinfo{booktitle}{\emph{Neural network design}}.
  Vol.~\bibinfo{volume}{20}.
\newblock \bibinfo{publisher}{Pws Pub. Boston}.
\newblock


\bibitem[\protect\citeauthoryear{Hayhurst}{Hayhurst}{2001}]%
        {hayhurst2001practical}
\bibfield{author}{\bibinfo{person}{Kelly Hayhurst}.}
  \bibinfo{year}{2001}\natexlab{}.
\newblock \bibinfo{booktitle}{\emph{A practical tutorial on modified
  condition/decision coverage}}. Vol.~\bibinfo{volume}{210876}.
\newblock \bibinfo{publisher}{DIANE Publishing}.
\newblock


\bibitem[\protect\citeauthoryear{Haykin et~al\mbox{.}}{Haykin
  et~al\mbox{.}}{2009}]%
        {haykin2009neural}
\bibfield{author}{\bibinfo{person}{Simon Haykin} {et~al\mbox{.}}}
  \bibinfo{year}{2009}\natexlab{}.
\newblock \bibinfo{booktitle}{\emph{Neural networks and learning machines/Simon
  Haykin.}}
\newblock \bibinfo{publisher}{New York: Prentice Hall,}.
\newblock


\bibitem[\protect\citeauthoryear{Huang, Kwiatkowska, Wang, and Wu}{Huang
  et~al\mbox{.}}{2017}]%
        {huang2017safety}
\bibfield{author}{\bibinfo{person}{Xiaowei Huang}, \bibinfo{person}{Marta
  Kwiatkowska}, \bibinfo{person}{Sen Wang}, {and} \bibinfo{person}{Min Wu}.}
  \bibinfo{year}{2017}\natexlab{}.
\newblock \showarticletitle{Safety verification of deep neural networks}. In
  \bibinfo{booktitle}{\emph{Computer Aided Verification}}. Springer,
  \bibinfo{pages}{3--29}.
\newblock


\bibitem[\protect\citeauthoryear{{IEEE}}{{IEEE}}{2008}]%
        {4610935}
\bibfield{author}{\bibinfo{person}{{IEEE}}.} \bibinfo{year}{2008}\natexlab{}.
\newblock \bibinfo{booktitle}{\emph{{IEEE} Standard For Floating-Point
  Arithmetic}}.
\newblock
\newblock
\shownote{{IEEE} 754-2008.}


\bibitem[\protect\citeauthoryear{Islam and Raj}{Islam and Raj}{2017}]%
        {Islam2017}
\bibfield{author}{\bibinfo{person}{Kh Islam} {and} \bibinfo{person}{Ram Raj}.}
  \bibinfo{year}{2017}\natexlab{}.
\newblock \showarticletitle{Real-Time (Vision-Based) Road Sign Recognition
  Using an Artificial Neural Network}.
\newblock \bibinfo{journal}{\emph{Sensors}} \bibinfo{volume}{17},
  \bibinfo{number}{4} (\bibinfo{year}{2017}).
\newblock
\showISSN{1424-8220}


\bibitem[\protect\citeauthoryear{Kahlon, Wang, and Gupta}{Kahlon
  et~al\mbox{.}}{2009}]%
        {KahlonWG09}
\bibfield{author}{\bibinfo{person}{Vineet Kahlon}, \bibinfo{person}{Chao Wang},
  {and} \bibinfo{person}{Aarti Gupta}.} \bibinfo{year}{2009}\natexlab{}.
\newblock \showarticletitle{Monotonic Partial Order Reduction: An Optimal
  Symbolic Partial Order Reduction Technique}. In
  \bibinfo{booktitle}{\emph{{C}omputer-{A}ided {V}erification}},
  Vol.~\bibinfo{volume}{5643}. \bibinfo{pages}{398--413}.
\newblock


\bibitem[\protect\citeauthoryear{Katz, Barrett, Dill, Julian, and
  Kochenderfer}{Katz et~al\mbox{.}}{2017a}]%
        {katz2017reluplex}
\bibfield{author}{\bibinfo{person}{Guy Katz}, \bibinfo{person}{Clark Barrett},
  \bibinfo{person}{David Dill}, \bibinfo{person}{Kyle Julian}, {and}
  \bibinfo{person}{Mykel Kochenderfer}.} \bibinfo{year}{2017}\natexlab{a}.
\newblock \showarticletitle{Reluplex: An efficient SMT solver for verifying
  deep neural networks}. In \bibinfo{booktitle}{\emph{Computer Aided
  Verification}}. Springer, \bibinfo{pages}{97--117}.
\newblock


\bibitem[\protect\citeauthoryear{Katz, Pasareanu, and Barrett}{Katz
  et~al\mbox{.}}{2017b}]%
        {DBLP:journals/corr/abs-1710-00486}
\bibfield{author}{\bibinfo{person}{Guy Katz}, \bibinfo{person}{Corina
  Pasareanu}, {and} \bibinfo{person}{Clark Barrett}.}
  \bibinfo{year}{2017}\natexlab{b}.
\newblock \showarticletitle{DeepSafe: A Data-driven Approach for Checking
  Adversarial Robustness in Neural Networks}.
\newblock  (\bibinfo{date}{10} \bibinfo{year}{2017}).
\newblock


\bibitem[\protect\citeauthoryear{Krichen and Tripakis}{Krichen and
  Tripakis}{2009}]%
        {KrichenT09}
\bibfield{author}{\bibinfo{person}{Moez Krichen} {and} \bibinfo{person}{Stavros
  Tripakis}.} \bibinfo{year}{2009}\natexlab{}.
\newblock \showarticletitle{Conformance testing for real-time systems}.
\newblock \bibinfo{journal}{\emph{Formal Methods in System Design}}
  \bibinfo{volume}{34}, \bibinfo{number}{3} (\bibinfo{year}{2009}),
  \bibinfo{pages}{238--304}.
\newblock


\bibitem[\protect\citeauthoryear{Kroening}{Kroening}{2018}]%
        {cprover-manual}
\bibfield{author}{\bibinfo{person}{Daniel Kroening}.}
  \bibinfo{year}{2018}\natexlab{}.
\newblock \bibinfo{title}{{CP}rover {M}anual}.
\newblock \bibinfo{howpublished}{\url{http://www.cprover.org/cprover-manual/}}.
\newblock
\newblock
\shownote{[Online; accessed September-2018].}


\bibitem[\protect\citeauthoryear{Laboratory}{Laboratory}{2020}]%
        {nnet-format}
\bibfield{author}{\bibinfo{person}{Stanford Intelligent~Systems Laboratory}.}
  \bibinfo{year}{2020}\natexlab{}.
\newblock \bibinfo{title}{{NN}et {R}epository}.
\newblock \bibinfo{howpublished}{\url{https://github.com/sisl/NNet}}.
\newblock
\newblock
\shownote{[Online; accessed August-2020].}


\bibitem[\protect\citeauthoryear{Li, Li, Sawaya, Gopalakrishnan, Ghosh, and
  Rajan}{Li et~al\mbox{.}}{2012}]%
        {LiLSGGR12}
\bibfield{author}{\bibinfo{person}{Guodong Li}, \bibinfo{person}{Peng Li},
  \bibinfo{person}{Geoffrey Sawaya}, \bibinfo{person}{Ganesh Gopalakrishnan},
  \bibinfo{person}{Indradeep Ghosh}, {and} \bibinfo{person}{Sreeranga Rajan}.}
  \bibinfo{year}{2012}\natexlab{}.
\newblock \showarticletitle{{GKLEE:} concolic verification and test generation
  for {GPUs}}. In \bibinfo{booktitle}{\emph{17th Symposium on Principles and
  Practice of Parallel Programming}}. \bibinfo{pages}{215--224}.
\newblock


\bibitem[\protect\citeauthoryear{Li and Yuan}{Li and Yuan}{2017}]%
        {li2017convergence}
\bibfield{author}{\bibinfo{person}{Yuanzhi Li} {and} \bibinfo{person}{Yang
  Yuan}.} \bibinfo{year}{2017}\natexlab{}.
\newblock \showarticletitle{Convergence analysis of two-layer neural networks
  with relu activation}. In \bibinfo{booktitle}{\emph{Advances in Neural
  Information Processing Systems}}. \bibinfo{pages}{597--607}.
\newblock


\bibitem[\protect\citeauthoryear{Lundberg and Lee}{Lundberg and Lee}{2017}]%
        {LundbergL17}
\bibfield{author}{\bibinfo{person}{Scott Lundberg} {and}
  \bibinfo{person}{{Su-In} Lee}.} \bibinfo{year}{2017}\natexlab{}.
\newblock \showarticletitle{A Unified Approach to Interpreting Model
  Predictions}.
\newblock In \bibinfo{booktitle}{\emph{Advances in Neural Information
  Processing Systems 30}}, \bibfield{editor}{\bibinfo{person}{I.~Guyon},
  \bibinfo{person}{U.~Luxburg}, \bibinfo{person}{S.~Bengio},
  \bibinfo{person}{H.~Wallach}, \bibinfo{person}{R.~Fergus},
  \bibinfo{person}{S.~Vishwanathan}, {and} \bibinfo{person}{R.~Garnett}}
  (Eds.). \bibinfo{publisher}{Curran Associates, Inc.},
  \bibinfo{pages}{4765--4774}.
\newblock


\bibitem[\protect\citeauthoryear{Monteiro, Alves, Silva, Ismail, Cordeiro, and
  de~Lima-Filho}{Monteiro et~al\mbox{.}}{2018}]%
        {monteiro2018esbmc}
\bibfield{author}{\bibinfo{person}{Felipe Monteiro}, \bibinfo{person}{Erickson
  Alves}, \bibinfo{person}{Isabela Silva}, \bibinfo{person}{Hussama Ismail},
  \bibinfo{person}{Lucas Cordeiro}, {and} \bibinfo{person}{Eddie de
  Lima-Filho}.} \bibinfo{year}{2018}\natexlab{}.
\newblock \showarticletitle{{ESBMC-GPU} A context-bounded model checking tool
  to verify CUDA programs}.
\newblock \bibinfo{journal}{\emph{Science of Computer Programming}}
  \bibinfo{volume}{152} (\bibinfo{year}{2018}), \bibinfo{pages}{63--69}.
\newblock


\bibitem[\protect\citeauthoryear{Monteiro, Garcia, Cordeiro, and
  de~Lima~Filho}{Monteiro et~al\mbox{.}}{2017}]%
        {MonteiroGCF17}
\bibfield{author}{\bibinfo{person}{Felipe~R. Monteiro},
  \bibinfo{person}{M{\'{a}}rio Garcia}, \bibinfo{person}{Lucas~C. Cordeiro},
  {and} \bibinfo{person}{Eddie~Batista de Lima~Filho}.}
  \bibinfo{year}{2017}\natexlab{}.
\newblock \showarticletitle{Bounded model checking of {C++} programs based on
  the Qt cross-platform framework}.
\newblock \bibinfo{journal}{\emph{Softw. Test. Verification Reliab.}}
  \bibinfo{volume}{27}, \bibinfo{number}{3} (\bibinfo{year}{2017}).
\newblock


\bibitem[\protect\citeauthoryear{Murthy, Das, and Islam}{Murthy
  et~al\mbox{.}}{2019}]%
        {Murthy2019}
\bibfield{author}{\bibinfo{person}{Abhishek Murthy}, \bibinfo{person}{Himel
  Das}, {and} \bibinfo{person}{Md.~Ariful Islam}.}
  \bibinfo{year}{2019}\natexlab{}.
\newblock \bibinfo{booktitle}{\emph{Robustness of Neural Networks to Parameter
  Quantization}}.
\newblock \bibinfo{publisher}{Springer International Publishing},
  \bibinfo{address}{Cham}, \bibinfo{pages}{146--161}.
\newblock


\bibitem[\protect\citeauthoryear{Nvidia}{Nvidia}{2008}]%
        {nvidia2008cublas}
\bibfield{author}{\bibinfo{person}{CUDA Nvidia}.}
  \bibinfo{year}{2008}\natexlab{}.
\newblock \showarticletitle{Cublas library}.
\newblock \bibinfo{journal}{\emph{NVIDIA Corporation, Santa Clara, California}}
  \bibinfo{volume}{15}, \bibinfo{number}{27} (\bibinfo{year}{2008}),
  \bibinfo{pages}{31}.
\newblock


\bibitem[\protect\citeauthoryear{Odena, Olsson, Andersen, and Goodfellow}{Odena
  et~al\mbox{.}}{2019}]%
        {pmlr-v97-odena19a}
\bibfield{author}{\bibinfo{person}{Augustus Odena}, \bibinfo{person}{Catherine
  Olsson}, \bibinfo{person}{David Andersen}, {and} \bibinfo{person}{Ian
  Goodfellow}.} \bibinfo{year}{2019}\natexlab{}.
\newblock \showarticletitle{{T}ensor{F}uzz: Debugging Neural Networks with
  Coverage-Guided Fuzzing}. In \bibinfo{booktitle}{\emph{36th International
  Conference on Machine Learning}}. \bibinfo{pages}{4901--4911}.
\newblock


\bibitem[\protect\citeauthoryear{Pereira, Albuquerque, da~Silva, Marques,
  Monteiro, Ferreira, and Cordeiro}{Pereira et~al\mbox{.}}{2017}]%
        {PereiraASMMFC17}
\bibfield{author}{\bibinfo{person}{Phillipe Pereira}, \bibinfo{person}{Higo
  Albuquerque}, \bibinfo{person}{Isabela da Silva}, \bibinfo{person}{Hendrio
  Marques}, \bibinfo{person}{Felipe Monteiro}, \bibinfo{person}{Ricardo
  Ferreira}, {and} \bibinfo{person}{Lucas Cordeiro}.}
  \bibinfo{year}{2017}\natexlab{}.
\newblock \showarticletitle{SMT-based context-bounded model checking for {CUDA}
  programs}.
\newblock \bibinfo{journal}{\emph{Concurrency and Computation: Practice and
  Experience}} \bibinfo{volume}{29}, \bibinfo{number}{22}
  (\bibinfo{year}{2017}).
\newblock


\bibitem[\protect\citeauthoryear{Rocha, Rocha, Ismail, Cordeiro, and
  Fischer}{Rocha et~al\mbox{.}}{2017}]%
        {RochaRIC017}
\bibfield{author}{\bibinfo{person}{Williame Rocha}, \bibinfo{person}{Herbert
  Rocha}, \bibinfo{person}{Hussama Ismail}, \bibinfo{person}{Lucas Cordeiro},
  {and} \bibinfo{person}{Bernd Fischer}.} \bibinfo{year}{2017}\natexlab{}.
\newblock \showarticletitle{{DepthK: A k-Induction} Verifier Based on Invariant
  Inference for {C} Programs - (Competition Contribution)}. In
  \bibinfo{booktitle}{\emph{26th International Conference on Tools and
  Algorithms for the Construction and Analysis of Systems}}.
  \bibinfo{pages}{360--364}.
\newblock


\bibitem[\protect\citeauthoryear{Sena, Bessa, Ramalho, Cordeiro, and Mota}{Sena
  et~al\mbox{.}}{2019}]%
        {Sena20}
\bibfield{author}{\bibinfo{person}{Luiz Sena}, \bibinfo{person}{Iury Bessa},
  \bibinfo{person}{Mikhail Ramalho}, \bibinfo{person}{Lucas Cordeiro}, {and}
  \bibinfo{person}{Edjard Mota}.} \bibinfo{year}{2019}\natexlab{}.
\newblock \showarticletitle{Incremental Bounded Model Checking of Artificial
  Neural Networks in {CUDA}}. In \bibinfo{booktitle}{\emph{IX Brazilian
  Symposium on Computing Systems Engineering}}.
\newblock


\bibitem[\protect\citeauthoryear{Sun, Wu, Ruan, Huang, Kwiatkowska, and
  Kroening}{Sun et~al\mbox{.}}{2018}]%
        {SunWRHKK18}
\bibfield{author}{\bibinfo{person}{Youcheng Sun}, \bibinfo{person}{Min Wu},
  \bibinfo{person}{Wenjie Ruan}, \bibinfo{person}{Xiaowei Huang},
  \bibinfo{person}{Marta Kwiatkowska}, {and} \bibinfo{person}{Daniel
  Kroening}.} \bibinfo{year}{2018}\natexlab{}.
\newblock \showarticletitle{Concolic testing for deep neural networks}. In
  \bibinfo{booktitle}{\emph{33rd International Conference on Automated Software
  Engineering}}. \bibinfo{pages}{109--119}.
\newblock


\bibitem[\protect\citeauthoryear{Taly, Converse, and Pasareanu}{Taly
  et~al\mbox{.}}{2019}]%
        {DBLP:journals/corr/abs-1904-13215}
\bibfield{author}{\bibinfo{person}{Ankur Taly}, \bibinfo{person}{Hayes
  Converse}, {and} \bibinfo{person}{Corina Pasareanu}.}
  \bibinfo{year}{2019}\natexlab{}.
\newblock \showarticletitle{Finding Invariants in Deep Neural Networks}.
\newblock  (\bibinfo{date}{04} \bibinfo{year}{2019}).
\newblock


\bibitem[\protect\citeauthoryear{Wang, Zhang, Pasareanu, and Khurshid}{Wang
  et~al\mbox{.}}{2018b}]%
        {abs-1807-10439}
\bibfield{author}{\bibinfo{person}{Kaiyuan Wang}, \bibinfo{person}{Mengshi
  Zhang}, \bibinfo{person}{Corina Pasareanu}, {and} \bibinfo{person}{Sarfraz
  Khurshid}.} \bibinfo{year}{2018}\natexlab{b}.
\newblock \showarticletitle{Symbolic Execution for Deep Neural Networks}.
\newblock  (\bibinfo{date}{07} \bibinfo{year}{2018}).
\newblock


\bibitem[\protect\citeauthoryear{Wang, Pei, Whitehouse, Yang, and Jana}{Wang
  et~al\mbox{.}}{2018a}]%
        {Wang2018}
\bibfield{author}{\bibinfo{person}{Shiqi Wang}, \bibinfo{person}{Kexin Pei},
  \bibinfo{person}{Justin Whitehouse}, \bibinfo{person}{Junfeng Yang}, {and}
  \bibinfo{person}{Suman Jana}.} \bibinfo{year}{2018}\natexlab{a}.
\newblock \showarticletitle{{E}fficient {F}ormal {S}afety {A}nalysis of
  {N}eural {N}etworks}. In \bibinfo{booktitle}{\emph{Proceedings of the 32nd
  International Conference on Neural Information Processing Systems}}
  (Montr\'{e}al, Canada). \bibinfo{publisher}{Curran Associates Inc.},
  \bibinfo{pages}{6369–6379}.
\newblock


\bibitem[\protect\citeauthoryear{Xiao, Rasul, and Vollgraf}{Xiao
  et~al\mbox{.}}{2017}]%
        {mnist}
\bibfield{author}{\bibinfo{person}{Han Xiao}, \bibinfo{person}{Kashif Rasul},
  {and} \bibinfo{person}{Roland Vollgraf}.} \bibinfo{year}{2017}\natexlab{}.
\newblock \showarticletitle{Fashion-MNIST: a Novel Image Dataset for
  Benchmarking Machine Learning Algorithms}.
\newblock  (\bibinfo{date}{08} \bibinfo{year}{2017}).
\newblock


\bibitem[\protect\citeauthoryear{Yamaguchi, Brain, Ryder, Imai, and
  Kawamura}{Yamaguchi et~al\mbox{.}}{2019}]%
        {YamaguchiBRIK19}
\bibfield{author}{\bibinfo{person}{Tomoya Yamaguchi}, \bibinfo{person}{Martin
  Brain}, \bibinfo{person}{Chirs Ryder}, \bibinfo{person}{Yosikazu Imai}, {and}
  \bibinfo{person}{Yoshiumi Kawamura}.} \bibinfo{year}{2019}\natexlab{}.
\newblock \showarticletitle{{A}pplication {O}f {A}bstract {I}nterpretation {T}o
  {T}he {A}utomotive {E}lectronic {C}ontrol {S}ystem}. In
  \bibinfo{booktitle}{\emph{20th International Conference on Verification,
  Model Checking, and Abstract Interpretation}}. \bibinfo{pages}{425--445}.
\newblock


\bibitem[\protect\citeauthoryear{Zheng, Rogers, Luo, Dwyer, and Siegel}{Zheng
  et~al\mbox{.}}{2015}]%
        {ZhengRLDS15}
\bibfield{author}{\bibinfo{person}{Manchun Zheng}, \bibinfo{person}{Michael
  Rogers}, \bibinfo{person}{Ziqing Luo}, \bibinfo{person}{Matthew Dwyer}, {and}
  \bibinfo{person}{Stephen Siegel}.} \bibinfo{year}{2015}\natexlab{}.
\newblock \showarticletitle{{CIVL:} Formal Verification of Parallel Programs}.
  In \bibinfo{booktitle}{\emph{30th International Conference on Automated
  Software Engineering}}. \bibinfo{pages}{830--835}.
\newblock


\bibitem[\protect\citeauthoryear{Zheng, Song, Leung, and Goodfellow}{Zheng
  et~al\mbox{.}}{2016}]%
        {zheng2016improving}
\bibfield{author}{\bibinfo{person}{Stephan Zheng}, \bibinfo{person}{Yang Song},
  \bibinfo{person}{Thomas Leung}, {and} \bibinfo{person}{Ian Goodfellow}.}
  \bibinfo{year}{2016}\natexlab{}.
\newblock \showarticletitle{Improving the robustness of deep neural networks
  via stability training}. In \bibinfo{booktitle}{\emph{Conference on Computer
  Vision and Pattern Recognition}}. \bibinfo{pages}{4480--4488}.
\newblock


\end{thebibliography}


\end{document}